\documentclass[prb,twocolumn,showpacs,amsmath,amssymb,superscriptaddress]{revtex4}
\pdfoutput=1

\usepackage{epsfig}
\usepackage{ graphicx } 
\usepackage{ bm } 
\usepackage{ color }
\usepackage{ulem} 

\newcommand{\ua}{\uparrow}
\newcommand{\da}{\downarrow}

\newcommand{\be}{\begin{equation}}
\newcommand{\ee}{\end{equation}}
\newcommand{\bea}{\begin{eqnarray}}
\newcommand{\eea}{\end{eqnarray}}

\newcommand{\bs}{\boldsymbol}

\definecolor{green}{rgb}{0,0.75,0.3}

\newcommand{\veps}{\varepsilon}
\newcommand{\vk}{{\boldsymbol k}}

\begin{document}
\title{ Entanglement modes and topological phase transitions in superconductors }
\author{ T. P. Oliveira }
\email{ tharnier@fisica.ufsc.br }
\author{ P. D. Sacramento }
\email{ pdss@cfif.ist.utl.pt }
\affiliation{ \textit Centro de F\'isica das Intera\c c\~oes Fundamentais,
Instituto Superior T\'ecnico, Universidade de Lisboa, Av. Rovisco Pais, 1049-001 Lisboa, Portugal }
\date{ \today }


\begin{abstract}
Topological insulators and topological superconductors (TSC)
display various topological phases that are characterized by different
Chern numbers or by gapless edge states. In this work we show that
various quantum information methods
such as the von Neumann entropy, entanglement spectrum, fidelity and fidelity spectrum,
may be used to detect and distinguish topological phases and their transitions.
As an example we consider a two-dimensional $p$-wave superconductor, with Rashba spin-orbit coupling and a Zeeman term.
The nature of the
phases and their changes are clarified by the eigenvectors of the $k$-space
reduced density matrix.
We show that in the topologically non-trivial phases the highest weight eigenvector is fully
aligned with the triplet pairing state.
A signature of the various phase transitions between two points on the
parameter space is encoded in the $k$-space fidelity operator.
\end{abstract}

\pacs{03.65.Ud, 03.67.Mn, 74.40.Kb, 03.65.Vf}

\maketitle

\section{Introduction}

Topological phases are non-trivial phases of matter characterized by global entanglement
and correlations.
Various examples have received increased attention such as the traditional
quantum Hall phases, spin liquids, topological insulators and topological superconductors.
Due to their nature, topological phases are robust to local perturbations and have
received wide attention as possible elements for error-free quantum computation.

In two dimensions, time reversal breaking $\mathbb{Z}$~topological insulators
exhibit a charge Hall conductivity that is quantized and
proportional to the Chern number of the occupied bands \cite{TKNN}.
Time reversal invariant $\mathbb{Z}_2$ topological insulators have also been proposed and
found experimentally \cite{Hasan,Zhang,Konig,Hsieh,Xia}.
Such nontrivial topological phases are also characterized
by the presence of gapless edge modes \cite{Halperin,Hatsugai}.

Superconductivity
with non-trivial topology may also be obtained \cite{Zhang}. It can be due to
the pairing symmetry, as is the case of $p$-wave SCs \cite{ReadGreen}.
In semiconductors with Rashba SO coupling it arises when $s$-wave superconductivity
is induced
and a Zeeman term is added \cite{SatoPRL09, SauPRL10}.
In case the normal phase is already topologically non-trivial, a
TSC may be obtained if $s$-wave superconductivity is induced by proximity effect \cite{Fu,QHZ10}.
Topological phases and their phase transitions in topological superconductors may be
detected through the existence of
zero bias peaks in tunneling
spectroscopy experiments \cite{Mourik,Das,Deng,Fan} due to
Majorana end states,
including spatially resolved
peaks \cite{Yazdani}, or through anomalous Fraunhofer patterns or fractional Josephson effects
\cite{Veldhorst,Williams,Rokhinson}. Multiple Andreev reflection current in voltage-biased
junctions has also been proposed as a signature of topological order \cite{Pablo}, in
particular, in the derivative of the current and in the zero-bias conductance of nodal
noncentrosymmetric SCs \cite{Schnyder}. The imaginary part of $\sigma_{xy}$
at finite frequency may also be used to signal topological phases \cite{Ojanen}, as well
as the Hall conductivity and its derivatives \cite{us3}.

On the other hand, the interplay between quantum information and condensed matter physics has been
extensively considered,
such as the use of entanglement \cite{amico} in the study of zero-temperature quantum phase
transitions \cite{sachdev}.
This interrelation has been explored in the reanalysis of several
non-trivial exactly solvable models, using different
information measures to better understand the underlying physics
\cite{latorre-2004-4,gu-2005-71,gu_pra_68,gu,goteborg,vidal-2006-73,vidal-2006,gu_prl_93}.
For a bipartite system,
besides the von Neumann entropy
and related quantities\cite{nielsen.chuang}, other information measures like the
concurrence\cite{wooters}, the mutual information\cite{vedral_njp6,gu_qp_1,gu_qp_2},
the negativity\cite{vidal}, or the Meyer-Wallach\cite{meyer} and the
generalized global entanglement measures\cite{oliveira_pra_73,miranda} have been considered.
The effect
of the quantum statistics has been analyzed for free electrons\cite{lunkes05-2}
and bosons\cite{heaney06}, and
for electrons in a BCS superconductor\cite{oh04}.

The distinguishability between states through the fidelity, has also been used as a possible criterion
to detect
quantum phase transitions.
By its own nature, fidelity between pure ground states signals a change of state as one approaches a quantum
phase transition \cite{Gu,zanardi-first,zanardi-free_fermion,buonsante-prl,oelkers, chen-excited,min, zhou,
zanardi-differential,zanardi-scaling,wen-long-thermal}.
The fidelity between mixed states has also been
used as a signature of quantum phase transitions \cite{Zhou,us} and to distinguish
between different states of matter at finite temperatures \cite{BCS, Zanardi_t1, Zanardi_t2}.

As argued in Ref. \onlinecite{Haldane}, a more detailed information about a mixed state
may be obtained if the entanglement spectrum is analyzed.
Considering reduced density matrices, where part of the degrees of freedom are
integrated over,
it was shown, in the context of the quantum Hall effect \cite{Haldane, Bernevig} and
in the context of coupled spin chains \cite{Poilblanc}, that the ground partitioning state entanglement
spectrum of a subsystem $A$ contains information about excited energy states, of the frontier
of the subsystem $A$ with the complementary subsystem $B$.
Other partitioning of the system have been proposed that lead
to further information \cite{Bernevig2}. Considering a partitioning in momentum
space, it was shown that information about energy excitations of a single Heisenberg
chain is contained in the groundstate wave function, through the entanglement spectrum
\cite{Arovas}.

One may also consider \cite{us2,guafter} the spectrum of the fidelity operator ${\cal F}(\rho_1,\rho_2)$
between two density matrices, $\rho_1$ and $\rho_2$.
Its set of eigenvalues $f_n$, which is denoted {\it fidelity operator spectrum},
and $-\ln f_n$, called the {\it fidelity spectrum}, provide more information
as compared to the fidelity (its trace). This parallels the extra information
provided by the entanglement spectrum \cite{Haldane}, as compared to the von Neumann entropy.
In the case of two equal mixed states,
the operator ${\cal F}$ has a set of eigenvalues, $f_n=\Lambda_n$, such that
$-\ln \Lambda_n$ reduces to the entanglement spectrum.

Quantum information methods have also been used to detect the more elusive
topological phases and transitions. Kosterlitz-Thouless transitions were
successfully detected calculating the fidelity susceptibility of the $XXZ$ spin
chain \cite{min,chen77}.
A study of the one-dimensional asymmetric Hubbard model showed that,
different scaling regimes of the fidelity susceptibility may be used to
distinguish the two phases \cite{gu77}.
Other systems that display topological phases were also studied, like spin-$1/2$ particles
on a torus \cite{hamma77},
the toric code model and the quantum eight-vertex model \cite{abasto78} and
the spin honeycomb Kitaev model \cite{yang78}.
In the case of the Hubbard model it was shown that the fidelity metric satisfies an
hyper scaling \cite{campus78} and in the Kitaev honeycomb model it was found a different
temperature scaling behavior in the different phases \cite{abasto79} and a divergent
fidelity per lattice site, showing that a local measure is also able to detect the
global entanglement regime \cite{zhao80}, as in other systems
\cite{eriksson,castelnovo,trebst,wang10}.

In this work we will be interested in applying various quantum information measures
to study the topological phases and phase transitions in a topological superconductor. In particular, we will
consider the entropy, entanglement spectrum, fidelity and fidelity spectrum, focusing
on global properties.
We consider a momentum space partitioning, since the Hamiltonian factorizes.
In general, the groundstate is a singlet and the entropy vanishes. At the transitions
between the various phases the spectrum is gapless and degenerate leading to a finite entropy.
The opposite result is obtained for the fidelity between two density matrices, defined close by
in parameter space; when both density matrices are defined in the same phase the fidelity
approaches one and when the transition is reached the fidelity has a minimum.
However, the momentum space contributions to both the entropy and
fidelity, or the entanglement spectrum and fidelity spectrum contain a richer structure.
Also, the eigenvectors of the reduced density matrix reveal the nature of the state in a given
phase and, in particular, the change across a topological transition from a trivial to a non-trivial phase.
A recent use of the fidelity and fidelity susceptibility was
carried out to study a one-dimensional spinless superconductor with end Majorana fermions \cite{wang2013}.
Here we consider a two-dimensional superconductor with a parity breaking Rashba spin-orbit
term that, in general, mixes singlet and triplet pairings \cite{Gorkov}. Also, a Zeemann term is added leading
to a rich phase diagram \cite{sato} with different topological phases, that can be characterized
by the Chern number, gapless edge states and winding number \cite{sato,us3}.

\begin{figure}[t]
\includegraphics[width=0.95\columnwidth]{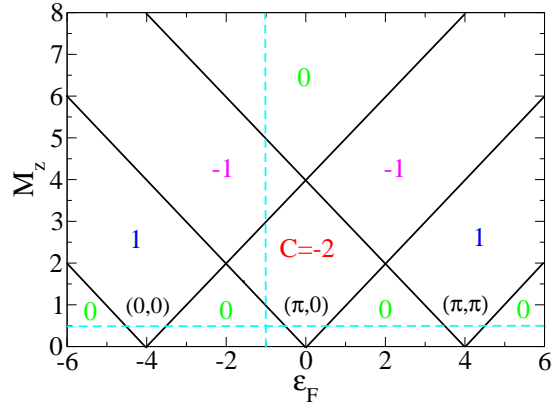}
\caption{\label{fig1}
(Color online) Topological phases and their Chern (C) numbers as a function of chemical potential and magnetization.
The transitions occur at three sets of momenta $\boldsymbol{k}=(0,0),\boldsymbol{k}=(\pi,0),\boldsymbol{k}=(\pi,\pi)$.}
\end{figure}

In section II we present the model studied and the entities used to detect the quantum transitions.
In section \ref{sec:von_Neumann} we present the calculation method and our results for the entropy.
In section \ref{subsec:entang_spectrum} we present results for
the entropy spectrum and the analysis of the band structure of the system.
Also, we show how the eingenvectors of the system can be
useful for a better understanding of the transitions.
The eigenvalues and eigenvectors of k-subspace reduced density matrices are considered in subsection
\ref{subsec:rentang_spectrum}.
In section \ref{sec:Fidelity} we present the calculation method and our results for the fidelity,
and show how it clearly detects the quantum transitions.
In section \ref{sec:conclusion} we present our main conclusions.
Finally, in the appendices we show in detail the results for the eigenvector of the largest eigenvalue
of the density matrix and the reduced density matrix.


\section{Topological superconductor}

We consider a two-dimensional triplet superconductor
with $p$-wave symmetry.
This model was studied in Refs. \onlinecite{sato,us3}.
We write the Hamiltonian as
\begin{eqnarray}
\hat H = \frac 1 2\sum_\vk  \left( {\boldsymbol c}_{\vk}^\dagger ,{\boldsymbol c}_{-\vk}   \right)
\left(\begin{array}{cc}
\hat H_0(\vk) & \hat \Delta(\vk) \\
\hat \Delta^{\dagger}(\vk) & -\hat H_0^T(-\vk) \end{array}\right)
\left( \begin{array}{c}
 {\boldsymbol c}_{\vk} \\  {\boldsymbol c}_{-\vk}^\dagger  \end{array}
\right)
\label{bdg1}
\end{eqnarray}
where $\left( {\boldsymbol c}_{\vk}^{\dagger}, {\boldsymbol c}_{-\vk} \right) =
\left( c_{\vk\ua}^{\dagger}, c_{\vk\da}^\dagger ,c_{-\vk\ua}, c_{-\vk\da}   \right)$
and
\begin{equation}
\hat H_0=\epsilon_\vk\sigma_0 -M_z\sigma_z + \hat H_R\,.
\end{equation}
Here, $\epsilon_{\boldsymbol{k}}=-2 t (\cos k_x + \cos k_y )-\veps_F$
is the kinetic part, $t$ denotes the hopping parameter set in
the following as the energy scale, $\veps_F$ is the
chemical potential,
$\boldsymbol{k}$ is a wave vector in the $xy$ plane, and we have taken
the lattice constant to be unity. Furthermore, $M_z$
is the Zeeman splitting term responsible for the magnetization,
in energy units.
The Rashba spin-orbit term is written as
\begin{equation}
\hat H_R = \boldsymbol{s} \cdot \boldsymbol{\sigma} = \alpha
\left( \sin k_y \sigma_x - \sin k_x \sigma_y \right)\,,
\end{equation}
 where
$\alpha$ is measured in the energy units
 and $\boldsymbol{s} =\alpha(\sin k_y,-\sin k_x, 0)$.
The matrices $\sigma_x,\sigma_y,\sigma_z$ are
the Pauli matrices acting on the spin sector, and $\sigma_0$ is the
$2\times2$ identity.

The pairing matrix reads
\begin{equation}
\hat \Delta = i\left( {\boldsymbol d}\cdot {\boldsymbol\sigma} + \Delta_s \right) \sigma_y =
 \left(\begin{array}{cc}
-d_x+i d_y & d_z + \Delta_s \\
d_z -\Delta_s & d_x +i d_y
\end{array}\right)\,.
\end{equation}
We consider a situation where the
spin-orbit coupling is such that the pairing is aligned \cite{Sigrist2} along the
spin-orbit vector $\boldsymbol{s}$
as $\boldsymbol{d}=(d_x,d_y,d_z) = ( d / \alpha ) \boldsymbol{s} $
and $d$ is a scale parameter.
This is a situation expected if the spin-orbit
is strong (other cases were considered in \cite{ahe,us3}).

The energy eigenvalues and eigenfunction may be obtained solving the Bogoliubov-de Gennes equations
\be
\label{bdg2}
\left(\begin{array}{cc}
\hat H_0(\vk) & \hat \Delta(\vk) \\
\hat \Delta^{\dagger}(\vk) & -\hat H_0^T(-\vk) \end{array}\right)
\left(\begin{array}{c}
u_n\\
v_n
\end{array}\right)
= \epsilon_{\boldsymbol{k},n}
\left(\begin{array}{c}
u_n\\
v_n
\end{array}\right).
\ee
The 4-component spinor can be written as
\be
\left(\begin{array}{c}
u_n\\
v_n
\end{array}\right)=
\left(\begin{array}{c}
u_n(\boldsymbol{k},\uparrow) \\
u_n(\boldsymbol{k},\downarrow) \\
v_n(-\boldsymbol{k},\uparrow) \\
v_n(-\boldsymbol{k},\downarrow) \\
\end{array}\right) .
\ee
The energy eigenvalues can be written as
\be
\epsilon_{\boldsymbol{k},n} = \alpha_1 \sqrt{z_1+2 \alpha_2 \sqrt{z_2}} ,
\label{bdgbands}
\ee
where $\alpha_{1,2}=\pm 1$ and
\be
z_1= \boldsymbol{d}\cdot \boldsymbol{d} + \boldsymbol{s}\cdot \boldsymbol{s} + \epsilon_{\boldsymbol{k}}^2 +M_z^2 + \Delta_s^2
\ee
and
\bea
z_2 &=& \epsilon_{\boldsymbol{k}}^2 \left( \boldsymbol{s} \cdot \boldsymbol{s} + M_z^2 \right) + \left[ \left(
\boldsymbol{d} \times \boldsymbol{s} \right)_z \right]^2
\nonumber \\
&+& \Delta_s \left( \Delta_s ( \boldsymbol{d} \cdot \boldsymbol{d} + M_z^2 ) +2 \epsilon_{\boldsymbol{k}}
\boldsymbol{d} \cdot \boldsymbol{s} \right) .
\eea
The gap of the lowest band closes when $z_1=2\sqrt{z_2}$.

Since a topological transition may occur when a gap closes, the gapless points
may indicate the presence of topological transitions.
One way to characterize various topological phases is through the Chern number,
obtainable as an integral over the Brillouin zone of the Berry curvature \cite{xiao,Fukui}.
The locations of the gapless points may also be detected by the Berry curvature
of the bands.
When two bands become close there is a peak in the Berry curvature.
It can be shown that for the lowest band,
there are sharp peaks at the time-reversal momenta (as defined in Ref. \onlinecite{sato})
$\boldsymbol{k}=(0,0)$, $\boldsymbol{k}=(\pi,0),(0,\pi)$ and $\boldsymbol{k}=(\pi,\pi)$.
The second band shows peaks at the various
characteristic momenta, as discussed above for the entropy.
Summing over the occupied bands the Chern number has been calculated \cite{sato,us3}.
The results in the parameter space are shown in Fig. \ref{fig1}
using the typical parameters $t=1$, $\alpha=0.6$, $d=0.6$ and $\Delta_s = 0.1$.
From now on all figures use the same set of parameters, unless stated otherwise.

Non-trivial topological order for non-interacting Hamiltonians can be
related with the presence or absence of time-reversal symmetry (TRS), particle-hole symmetry
and chiral symmetry~\cite{Ludwig,LudwigAIP,LudwigNJP}. In Bogoliubov-de Gennes systems, particle-hole symmetry
is always present, and TRS is determinant to the nature of possible
topological phases in two dimensions. The superconductor we consider here
is time-reversal invariant if the Zeeman term is absent.
The system then belongs to the symmetry class DIII where the topological invariant is
a $\mathbb{Z}_2$ index
(the TRS operator $\mathcal{T}$ is such that $\mathcal{T}^2 = -1 $).
If the Zeeman term is finite, TRS is broken and the system belongs
to the symmetry class D
(the TRS operator $\mathcal{T}$ is such that $\mathcal{T}^2 = 0$).
The topological invariant that characterizes this phase is the first Chern number $C$,
and the system is said to be a $\mathbb{Z}$~topological superconductor.

Due to the bulk-edge correspondence, complementary information on the topological phases and
transitions may be obtained analyzing the edge states.
If $M_z=0$ and the pairing is $s$-wave,
the system is in a topologically trivial phase:
there is only the bulk gap and no gapless (edge) states.
In the case of p-wave or when
there is a mixture of s- and p-wave components, and
the amplitude of the p-wave pairing is larger than the corresponding
amplitude of the s-wave case, there are
two counter-propagating edge modes that
give opposite contributions to the total Chern number, $C=0$ ($\mathbb{Z}_2$ phase) \cite{sato}.

As the magnetization is turned on, TRS is broken.
For small magnetization, the superconductor is in a trivial phase
with Chern number $C=0$. A finite magnetization is then necessary to cause a topological
phase transition to a phase with non-zero Chern number~\cite{sato}.
This happens both for the
p-wave case and the s-wave case.
The sequence of Chern numbers is clearly correlated with the number of pairs
of edge states as shown in Ref. \onlinecite{sato}.
It is interesting to note that even though the system is in a $C=0$ phase the
number of edge states is two; the same as that in the parent $\mathbb{Z}_2$ phase,
when $M_z = 0$~\cite{sato}.
The presence of edge modes induced by bulk topology can also be shown
using dimensional reduction and thereby calculating the winding number~\cite{Wen}.
The calculation of the winding number gives the number of gapless edge modes both when the Chern number vanishes
and when the Chern number is finite~\cite{sato}.

\begin{figure}[h]
\includegraphics[width=0.98\columnwidth]{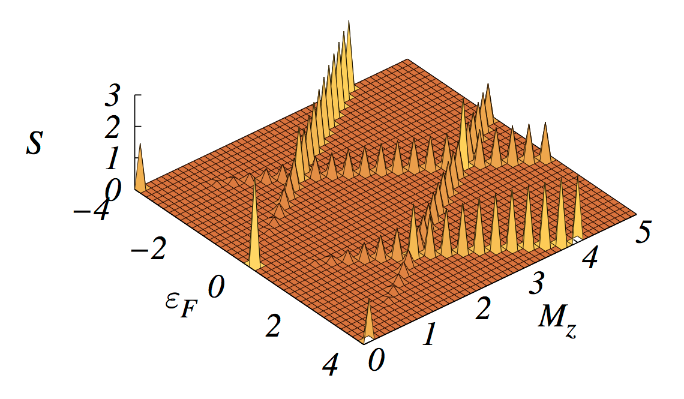}
\caption{\label{fig2}
(Color online) Total entropy as a function of chemical potential and magnetization,
for the typical parameters $t=1$, $\alpha=0.6$, $d=0.6$, $\Delta_s = 0.1$ and temperature
$T= 10^{-3}$.
}
\end{figure}


\section{von Neumann entropy}
\label{sec:von_Neumann}

Since the Hamiltonian is separable in momentum space,
the density matrix operator for a momentum $k$ may be defined as usual as
\begin{equation}
\label{ }
	\hat{\rho}_{k}	= \frac { \mathrm{e}^{-\beta \widehat{H}_{k} }  } { Z_{k} },
\end{equation}
with $ \widehat{H} = (1/2) \sum_{k} \widehat{H}_{k} $, defined from equation (\ref{bdg1}).
In the diagonal basis it is written as
\begin{equation}
\label{ }
	\rho_{k}=\left < n | \hat{\rho}_{k} | n \right > =
  	\frac { \mathrm{e}^{-\beta \left< n | \widehat{H}_{k} | n \right >  }  } { Z_{k} } .
\end{equation}

It is convenient to introduce a basis representation for the density matrix in terms of
the occupation numbers for a given momentum (and its symmetric) and the two spin
projections.
The eigenvalues of the density matrix are obtained if we diagonalize the Hamiltonian in the same basis.
We consider the representation
\begin{equation}
\label{eq:hamiltonian-new-base-state}
	\widetilde{H}_k	=\left < n_{k_{\uparrow}}  n_{- k_{\uparrow}}
   n_{k_{\downarrow}}  n_{-k_{\downarrow}} \right |
	\widehat{H}_{k} \left | n_{k_{\uparrow}}  n_{- k_{\uparrow}}  n_{k_{\downarrow}}  n_{-k_{\downarrow}} \right >
\end{equation}
The diagonalization of the Hamiltonian matrix in this enlarged basis is written as
\begin{equation}
\label{ }
	\widetilde{H}_k \bm{Q}_{k,n}=\lambda_{k,n} \bm{Q}_{k,n} \quad ; \quad n = 1, \ldots, 16
\end{equation}
note that $n$ here is just an index number and should not be confused with the occupation number of equation (\ref{eq:hamiltonian-new-base-state}).

In the same basis the density matrix may be written as
\begin{equation}
\label{ }
	\rho_{k}=\frac { \mathrm{e}^{-\beta \widetilde{H}_{k} }  } { Z_{k} } .
\end{equation}
Therefore the eigenvalues of the density matrix may be written as
$\rho_k \bm{Q}_{k,n} = \Lambda_{k,n} \bm{Q}_{k,n}$
where
\begin{equation}
\label{eq:eigval-DM}
	\Lambda_{k,n} =	\frac { \mathrm{e}^{ -\beta \lambda_{k,n} } } { \sum\limits_{n'} \mathrm{e}^{ -\beta \lambda_{k,n'} } } .
\end{equation}

The entropy for each momentum can be obtained using that
\begin{equation}
\label{ }
	S_{k}=- \sum\limits_{n=1}^{16} \Lambda_{k,n} \ln ( \Lambda_{k,n} )
\end{equation}
and the total entropy is obtained summing over the Brillouin zone
\begin{equation}
\label{ }
	S=\sum\limits_{k} S_{k} .
\end{equation}

\begin{figure}[h]
\includegraphics[width=0.9\columnwidth]{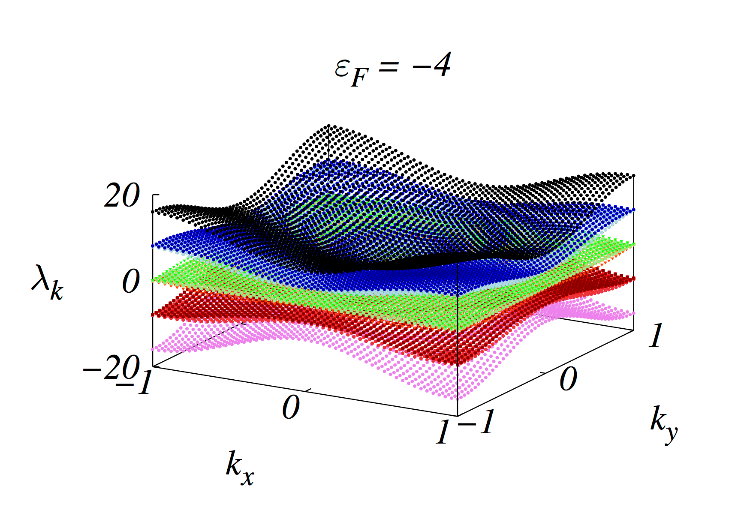}
\includegraphics[width=0.9\columnwidth]{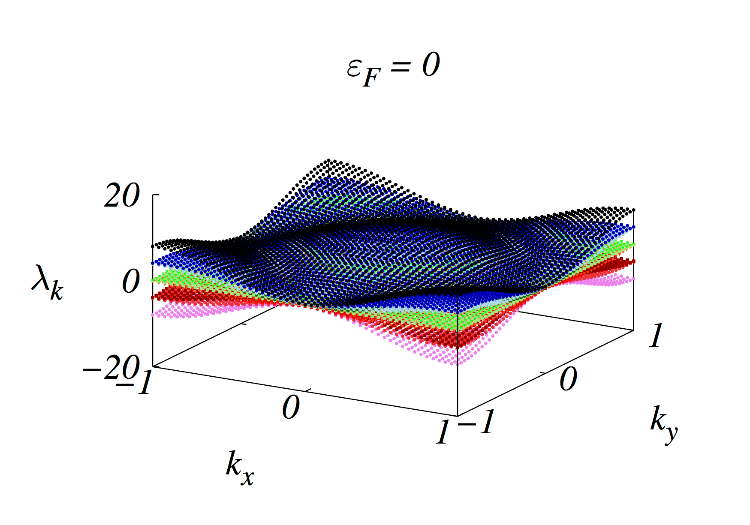}
\includegraphics[width=0.9\columnwidth]{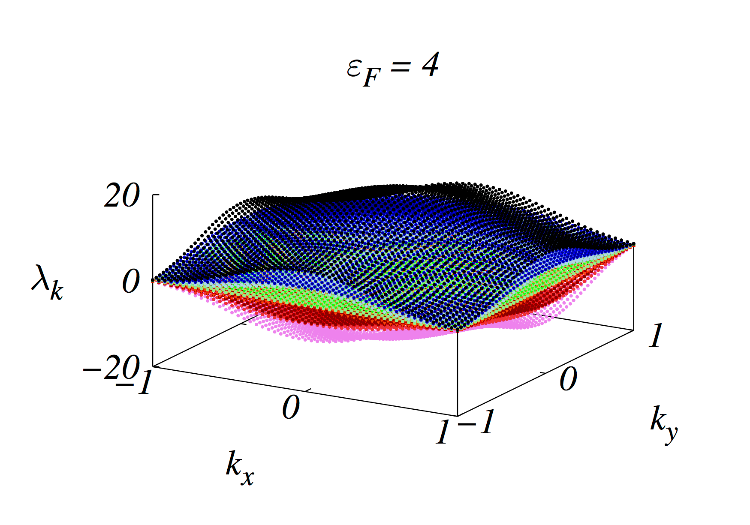}
\caption{\label{fig3}
(Color online) Bands of $\widetilde{H}_k$ for $M_z=0.1$ and $\veps_F = -4,0,4$, respectively.
}
\end{figure}

\begin{figure}[t]
\includegraphics[width=0.7\columnwidth]{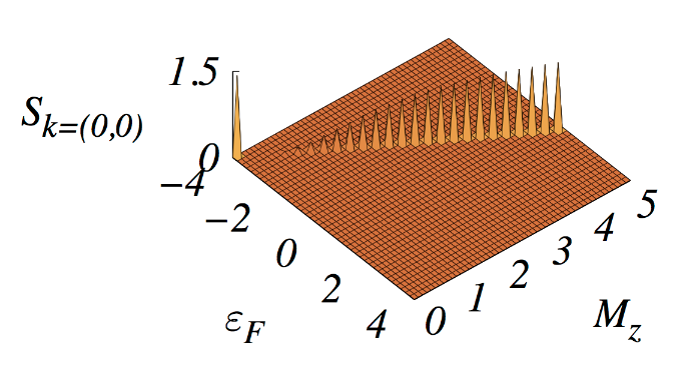}
\includegraphics[width=0.7\columnwidth]{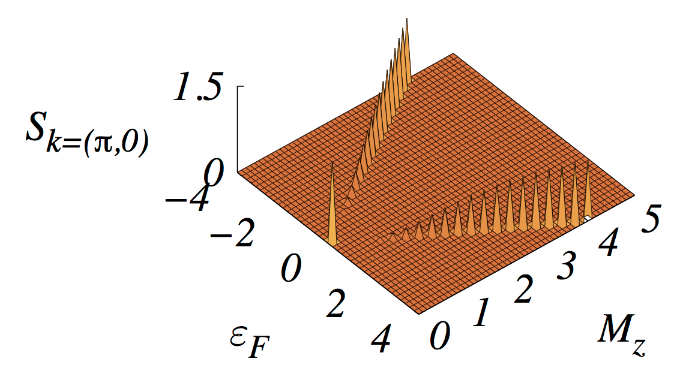}
\includegraphics[width=0.7\columnwidth]{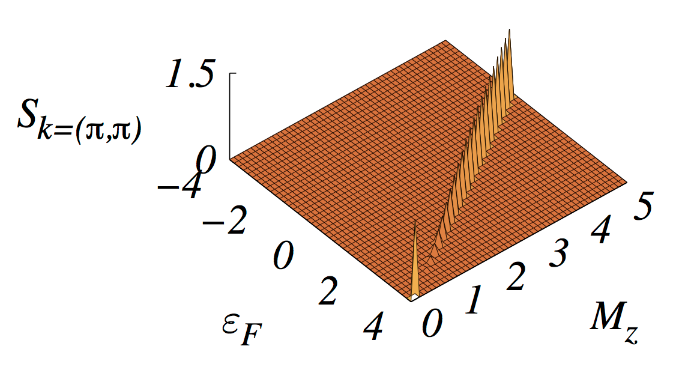}
\caption{\label{fig4}
(Color online) Entropy as a function of chemical potential and magnetization for $\mathbf{k}=(0,0), \mathbf{k}=(\pi,0),
\mathbf{k}=(\pi,\pi)$, respectively.
}
\end{figure}

The results for the entropy as a function of chemical potential and magnetization (Zeeman term)
are shown in Fig. \ref{fig2},
for a temperature parameter $T = 10^{-3}$, where $\beta = 1/T$ ($k_B=1$).
At very low temperature the entropy is of the type
$S=\ln \Omega$, where $\Omega$ is the degeneracy of the lowest state. In general, the lowest
state is non-degenerate and the entropy vanishes. At the transition points the system becomes
gapless and, at some points in momentum space, the lowest state becomes degenerate and the entropy
is finite. Integrating over momentum we sum over the various gapless points. As can be seen
comparing Fig. \ref{fig1} and Fig. \ref{fig2} the entropy tracks very well the topological
phase transitions. Except at the transitions, the entropy is featureless and does not distinguish
between the various topological phases.

Further insight is obtained if we analise the entropy
$S_k$ as a function of momentum, in order to understand the origin
of the peaks in Fig. \ref{fig2}. At $M_z=0.1$ there are three peaks in Fig. \ref{fig2} at chemical
potentials $\veps_F=-4,0,4$.
Those are special points, degenerate at absolute zero temperature, with values
$\ln (4)$, $2 \ln (4)$ and $\ln (4)$, respectively.
The fact that those points occur at $M_z = 0.1$ has to do with the choice of $\Delta_s = 0.1$
(for $\Delta_s=0$ those peaks occur at $M_z=0$).
These peaks
are due to singularities at momenta $\boldsymbol{k}=(0,0), \boldsymbol{k}=(\pi,0), \boldsymbol{k}=(\pi,\pi)$, respectively,
and their equivalent points in the Brillouin zone.
At very low temperature the peaks get very sharp. Therefore, a small finite temperature is used for
better visualization.

The energy spectrum degenerate gapless points are also seen in the bands of the matrix representation of the hamiltonian
$\widetilde{H}_k$.
These are shown in Fig. \ref{fig3}, where we have considered the same special points, $M_z = 0.1$, in the phase diagram.
Their relevance also extends beyond those particular chemical potential values. As shown in Fig. \ref{fig4}
the momenta responsible for the various transition lines is the same set. The lines that emerge
from $\veps_F=-4$ are associated with $\boldsymbol{k}=(0,0)$, the lines that emerge from $\veps_F=0$ are
associated with $\boldsymbol{k}=(\pi,0)$ and those from $\veps_F=4$ with $\boldsymbol{k}=(\pi,\pi)$.

\begin{figure}[t]
\includegraphics[width=0.95\columnwidth]{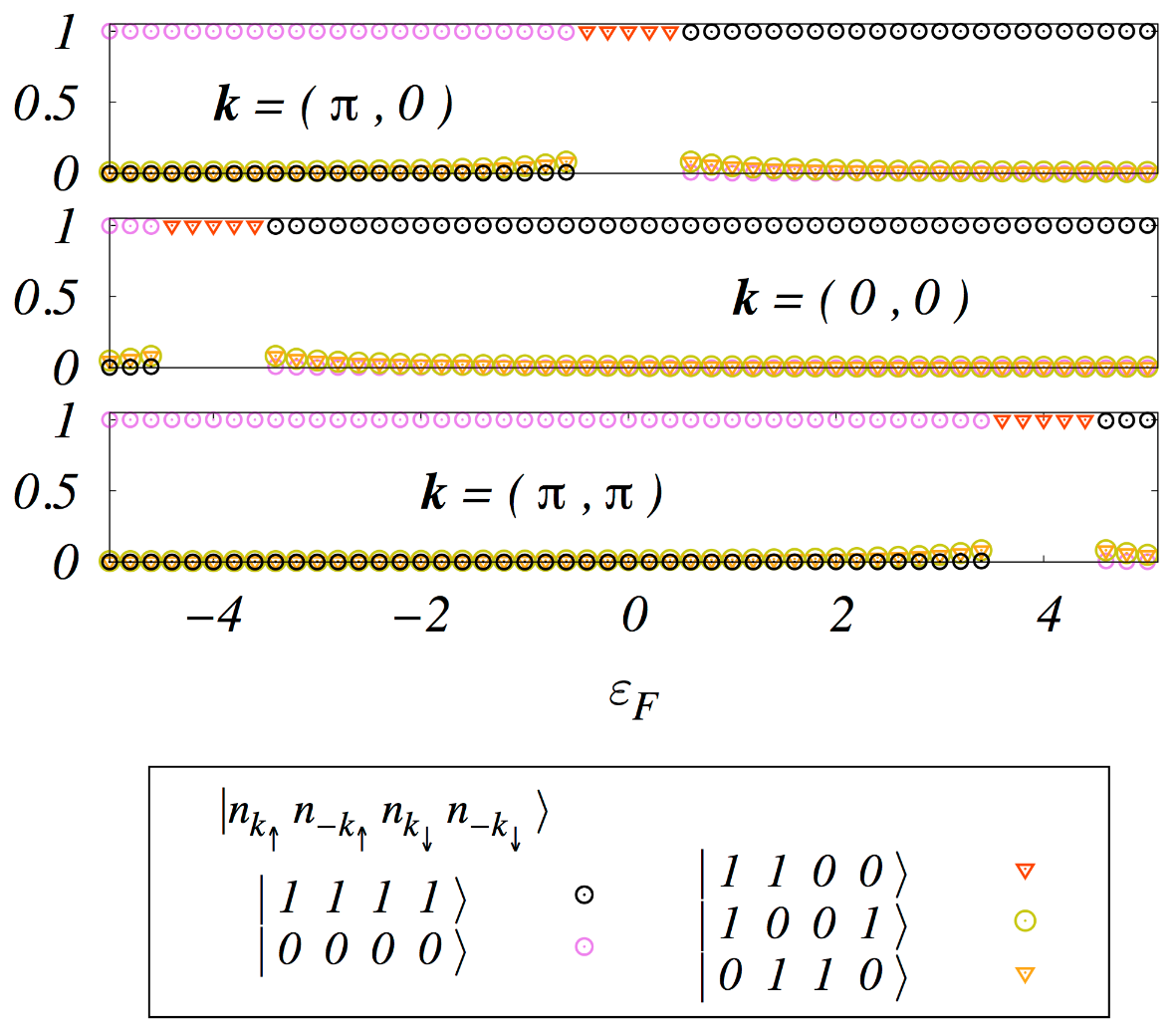}
\caption{\label{fig5}
(Color online) Eigenvector components for the highest eigenvalue of the reduced
density matrix $\rho_k$ for $M_z=0.5$ as a function of the chemical potential
$\veps_F$ for different momentum values
$\mathbf{k}=(0,0), \mathbf{k}=(\pi,0),
\mathbf{k}=(\pi,\pi)$.
The components that are not shown are nearly zero.
}
\end{figure}

\begin{figure}[t]
\includegraphics[width=0.95\columnwidth]{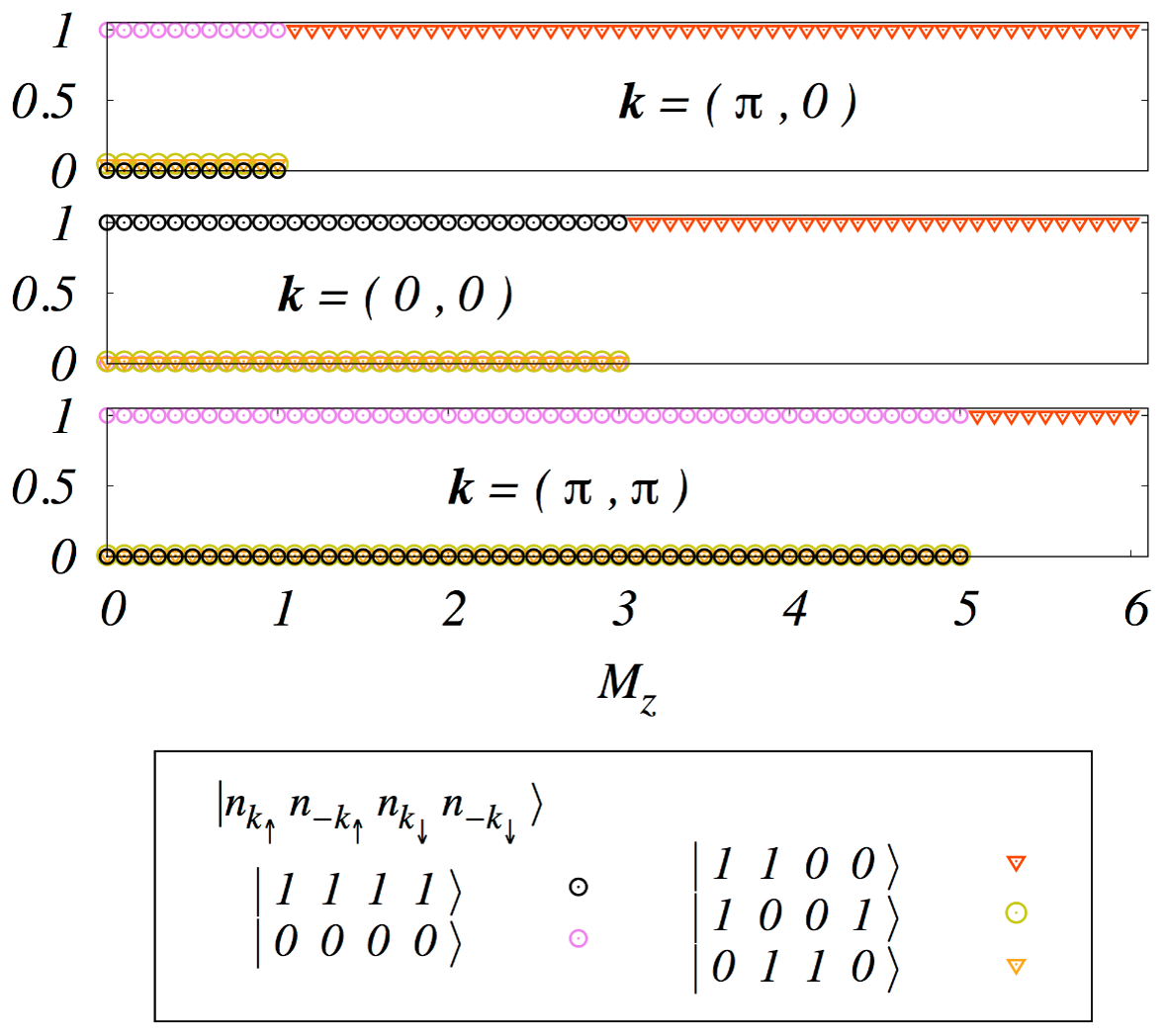}
\caption{\label{fig6}
(Color online) Eigenvector components for the highest eigenvalue of the reduced
density matrix $\rho_k$ for $\veps_F=-1$ as a function of the magnetization for  different momentum values
$\mathbf{k}=(0,0), \mathbf{k}=(\pi,0),
\mathbf{k}=(\pi,\pi)$.
The components that are not shown are nearly zero.
}
\end{figure}

\begin{figure}[h]
\includegraphics[width=0.9\columnwidth]{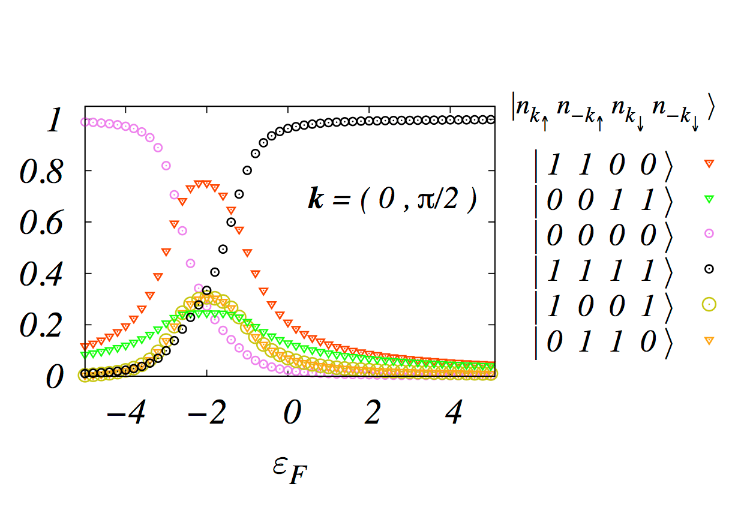}
\includegraphics[width=0.85\columnwidth]{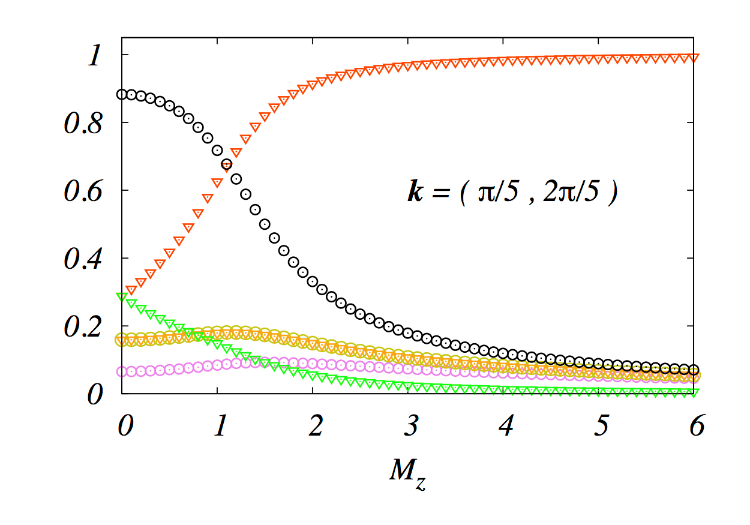}
\caption{\label{fig7}
(Color online) Eigenvector components of largest eigenvalue of reduced density matrix
$\rho_k$ for arbitrary momentum values for $M_z=0.5$ as a function of chemical potential (top panel)
and for $\veps_F=-1$ as a function of magnetization (lower panel).
The components that are not shown are nearly zero.
}
\end{figure}


\section{Entanglement spectrum and eigenvectors}
\label{subsec:entang_spectrum}

The entanglement spectrum of a bipartite system with a real space partition, provides interesting information
about the states along the interface between the two subsystems, as discussed in the
introduction. The momentum space partition carried out in this work, originating in
the Fourier partial diagonalization of the Hamiltonian, provides information about
the spectrum in each momentum separately.
Even though this partition is trivial, a detailed analysis reveals interesting information
about the entanglement of the system, through the entanglement spectrum and the eigenvectors
of the reduced density matrix.
The entanglement spectrum can be seen from Fig. \ref{fig3}
since it is simply the spectrum of the hamiltonian matrix $\widetilde{H}_k$, once both are diagonal in the same basis states.
At low temperatures, this translates to a spectrum of the
$k$-space density matrix where the lowest band has an eigenvalue close to unity and
all others close to zero (except at the degenerate points where the lowest state has
an eigenvalue which is the inverse of the degeneracy).

The eigenvectors provide
additional information about the phases of the system, and in particular provide interesting
information about the transitions.
In general, each eigenvector is written as a linear combination of
the 16 basis states used, $| n_{k_{\uparrow}}  n_{- k_{\uparrow}}  n_{k_{\downarrow}}  n_{-k_{\downarrow}} \rangle $.
It turns out that in most cases, out of the 16 coefficients, only a few are non-zero,
for each eigenvector. Their physical interpretation is in several cases clear, and it is possible to
understand the physical content of the dominant eigenstates, and what changes as a transition occurs.

The dominant states are: i) the states $|1 1 0 0 \rangle$ and $|0 0 1 1 \rangle$ corresponding to the spin triplet pairing,
ii) the state $|0 0 0 0 \rangle$ corresponding to an empty state in momentum and spin
spaces, iii) the state $|1 1 1 1 \rangle$ corresponding to a basis state fully occupied with 4 electrons
(for a given momentum and its symmetric and both spin components), iv) the states $|1 0 0 1 \rangle , |0 1 1 0 \rangle$
corresponding to the spin singlet pairing. Other states are also present but their coefficients are typically
quite small.

In Figs. \ref{fig5},\ref{fig6},\ref{fig7} we show results for the absolute value of the coefficients of the most important
contributions to the eigenvector, corresponding to the highest eigenvalue of the
$k$- density matrix (corresponding to the lowest band of the hamiltonian matrix).
In Fig. \ref{fig5} the magnetization is fixed, the chemical potential is changing and each panel corresponds
to the three singular momenta values. The set of chemical potential values are those corresponding to the horizontal
dashed line in Fig. \ref{fig1}. In Fig. \ref{fig6} the chemical potential is fixed at $\veps_F=-1$ and the
magnetization is varied tracing the vertical dashed line of Fig. \ref{fig1}. In Fig. \ref{fig7} we consider similar
plots but for arbitrary momenta values not corresponding to any of the singular momenta.

Along the dashed horizontal line the Chern number changes from $C=0 \rightarrow C=1\rightarrow C=0 \rightarrow C=-2
\rightarrow C=0 \rightarrow C=1 \rightarrow C=0$. The symmetry in the sequence of the Chern numbers
around half-filling ($\veps_F=0$) is however not directly manifested, neither in the results for the coefficients,
nor for the momenta where the transitions are detected.
But the symmetry comes out if we group the coefficients by similar spin pairing states, as shown in the appendix.
The transitions crossed by the dashed line are detected considering
$\boldsymbol{k}=(0,0), \boldsymbol{k}=(0,0), \boldsymbol{k}=(\pi,0), \boldsymbol{k}=(\pi,0), \boldsymbol{k}=(\pi,\pi),
\boldsymbol{k}=(\pi,\pi)$, respectively.
This is understood from the results of Fig. \ref{fig4}.

The most important contribution of the coefficients analysis is that they reveal the nature of the states in each phase.
An individual special $\boldsymbol{k}$ analysis shows that, for small chemical potential values, the empty state $|0 0 0 0 \rangle$ has the highest coefficient.
As a transition occurs, the spin triplet basis state $|1 1 0 0 \rangle$ becomes the dominant until the next transition occurs,
when the fully occupied state $|1 1 1 1 \rangle$ becomes the dominant one. This sequence happens at
different chemical potentials: those corresponding to the specific transition line for which, at given momentum
value, the lowest band becomes gapless. Close to the various transitions the spin singlet states $|1 0 0 1 \rangle $ and $|0 1 1 0 \rangle$ also contribute.

Similar results are obtained crossing the transitions by the vertical dashed line, shown in Fig. \ref{fig6}, where the sequence of transitions
is now observed as a function of varying magnetization, at fixed chemical potential.
However, it is shown that, after a given transition, the spin triplet state becomes the dominant one and prevails until
large values of the magnetization. Also, for low $M_z$ and before the transition, depending on the momenta considered, the dominant state
is either the empty state or the fully occupied state. It turns out that, counter-intuitively, for chemical potential
$\veps_F=-1$ the dominant state is the fully occupied state
for momenta close to the origin of the Brillouin zone, while the empty state is the dominant contribution close to the edges
of the Brillouin zone.

The sharpness of the contribution
of the various coefficients is lost, as shown in Fig. \ref{fig7}, if the momenta are not the singular points.
In this case no transitions are observed but their effect is still seen in broad features.

\begin{figure}[t]
\includegraphics[width=0.49\columnwidth]{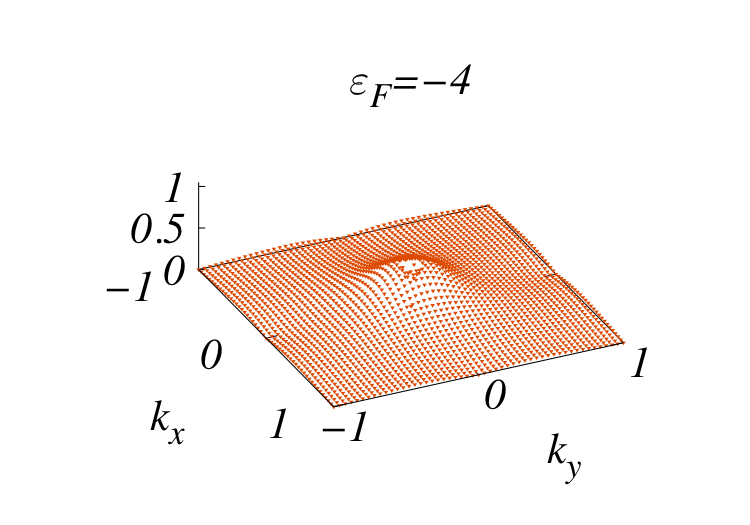}
\includegraphics[width=0.49\columnwidth]{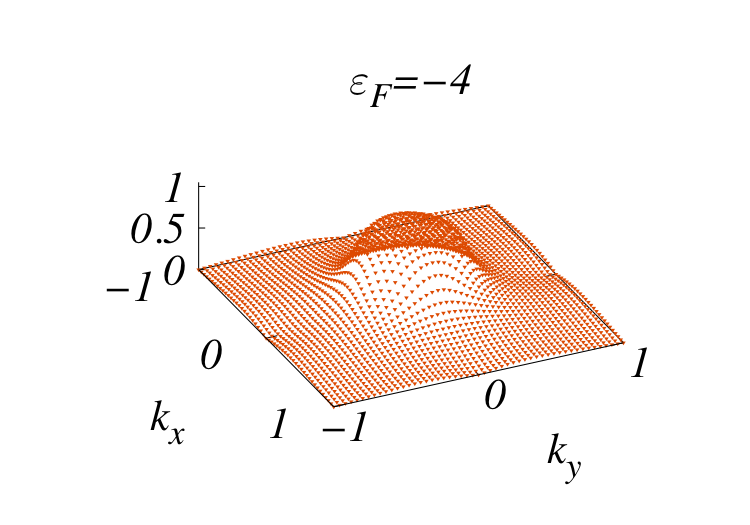}
\includegraphics[width=0.49\columnwidth]{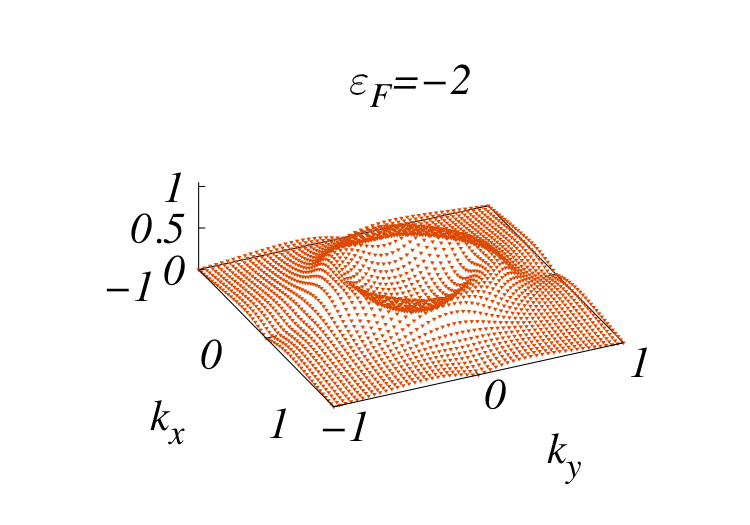}
\includegraphics[width=0.49\columnwidth]{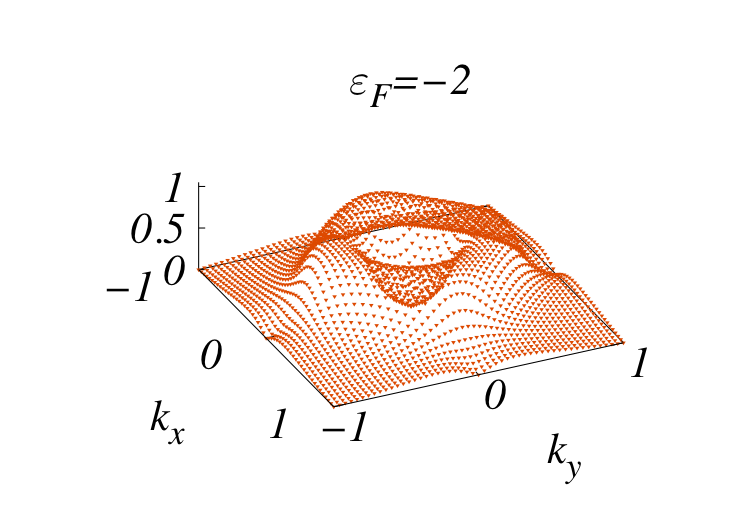}
\includegraphics[width=0.49\columnwidth]{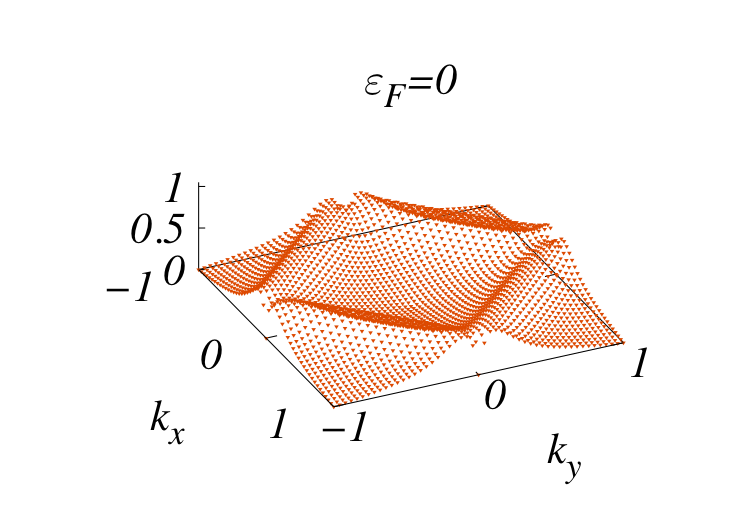}
\includegraphics[width=0.49\columnwidth]{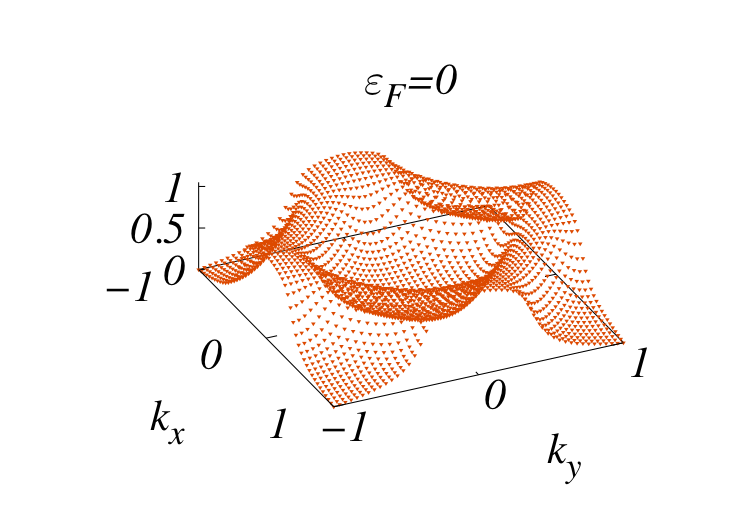}
\includegraphics[width=0.49\columnwidth]{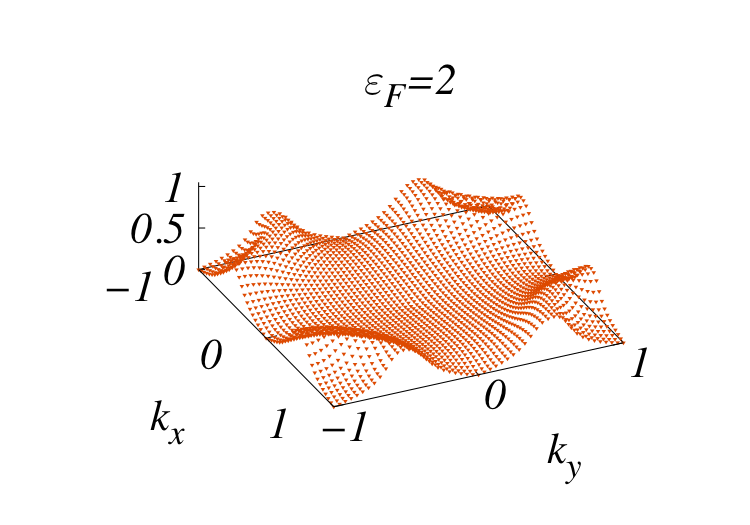}
\includegraphics[width=0.49\columnwidth]{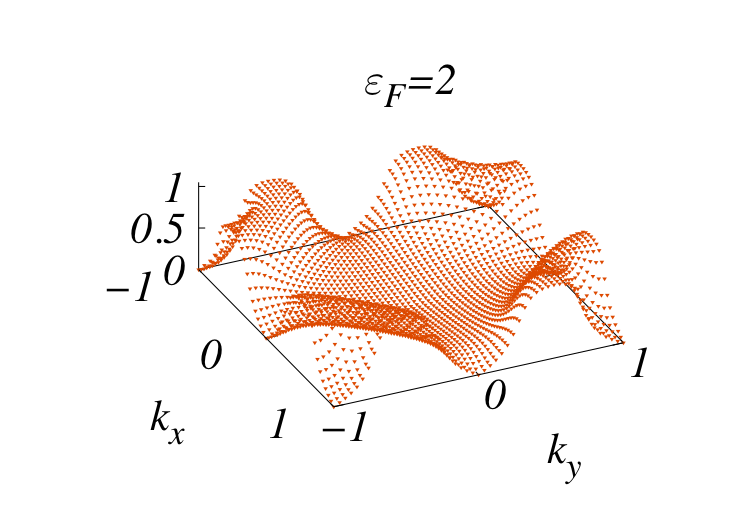}
\includegraphics[width=0.49\columnwidth]{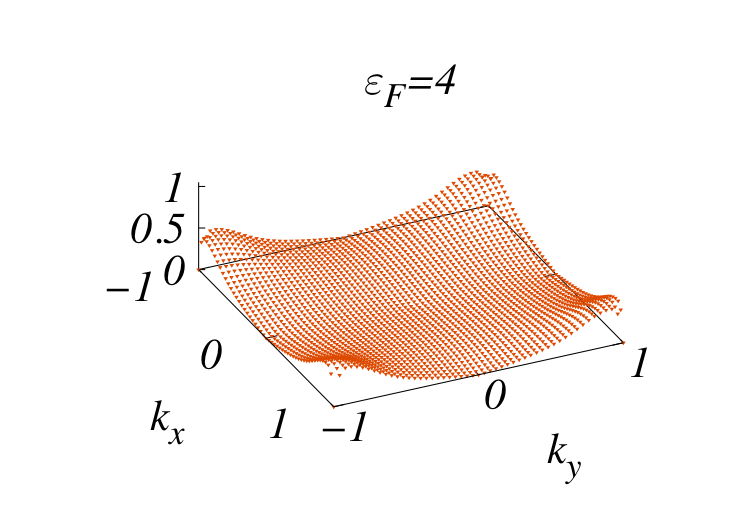}
\includegraphics[width=0.49\columnwidth]{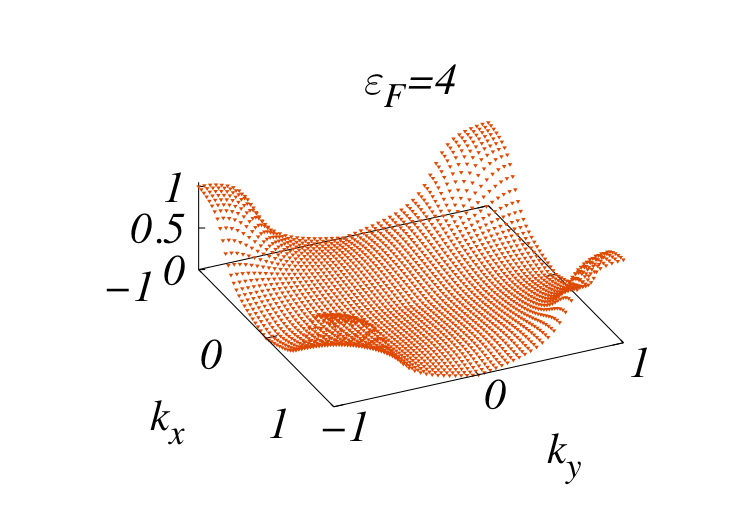}
\caption{\label{fig12}
(Color online) Absolute value of the coefficient of the spin triplet state as a function of momentum.
In the left column $M_z=0$ and in the right column $M_z=1$. The rows are for
$\veps_F=-4,-2,0,2,4$, respectively from top to bottom.
}
\end{figure}

As Figs. \ref{fig5}, \ref{fig6} indicate, the topological phases with finite Chern number, are associated
with a large coefficient of the triplet contribution. In Fig. \ref{fig12} we show the momentum dependence
of the absolute value of the triplet pairing coefficient for $M_z=0$ (left column) and $M_z=1$ (right column).
Note that for $M_z=0$ the system is in a $\mathbb{Z}_2$ phase in the region where $|\veps_F|<4$. For finite magnetization
the system is in a $\mathbb{Z}$ phase with either finite or vanishing Chern number.
For both values of the magnetization, there is a clear correlation between the absolute value of the
triplet coefficient and the location of the Fermi surface which expands as the chemical potential is
increased. At small chemical potential the Fermi surface is centered around zero momentum. Increasing
the chemical potential the Fermi surface will approach the point $\boldsymbol{k}=(\pi,0)$ and equivalent,
and as the chemical potential increases further the Fermi surface will approach the point
$\boldsymbol{k}=(\pi,\pi)$ and equivalents. The amplitude of the coefficient is larger in the case
of $M_z=1$. For $M_z=0$ the two spin orientations of the triplet state are degenerate and it can be
seen that the amplitude is fairly constant along the Fermi surface. When the magnetization is finite,
there is a larger amplitude in the neighborhood of the singular points previously found for the entanglement
entropy and fidelity. In this case, near these points, and depending on the chemical potential, the
coefficient is close to saturation, which implies that the triplet state is dominant over all other basis
states.

In Appendix A further detailed results for the eigenvector structure are presented. The results show that the
topological phases may be identified by the saturation of the spin triplet contribution.
The empty and doubly occupied states usually dominate over the Brillouin zone, except near the Fermi surface,
where the main contribution comes from the triplet and singlet pairing states.

\subsection{$k$-subspace reduced density matrix}
\label{subsec:rentang_spectrum}

\begin{figure}[t]
\includegraphics[width=0.6\columnwidth]{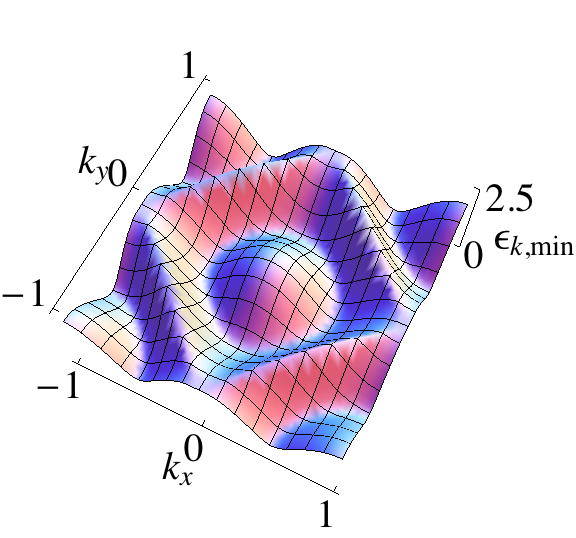}
\includegraphics[width=0.6\columnwidth]{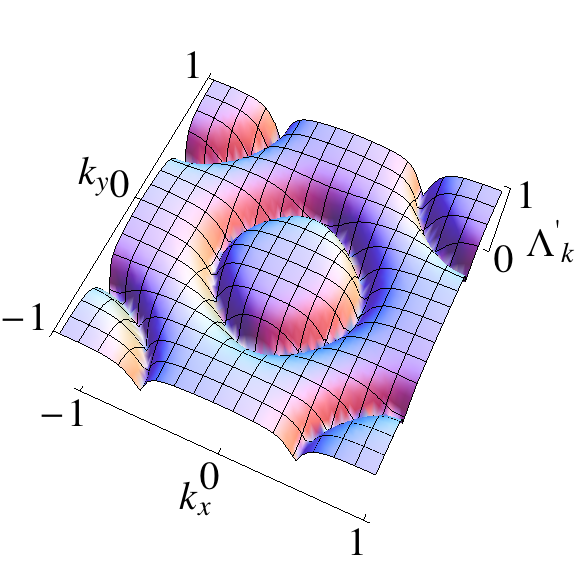}
\includegraphics[width=0.6\columnwidth]{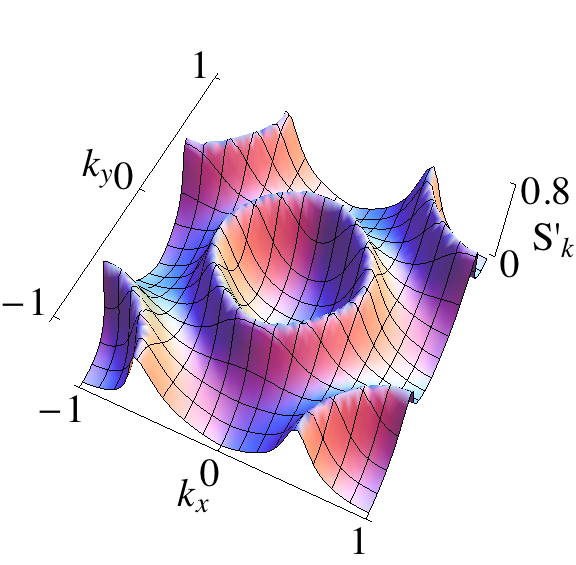}
\caption{\label{fig81}
(Color online) a) Lowest energy band, b) largest eigenvalue of reduced density matrix and
c) entropy as a function of momentum for $M_z=2, \veps_F=0$ ($C=-2$).
}
\end{figure}

Since the Hamiltonian is additive in momentum space the
density matrix is factorizable leading to a density matrix as expressed in
eq. (11).
We have diagonalized using the number occupation basis shown in the
same equation. This leads to a $16 \times 16$ matrix both for the
Hamiltonian and the density matrix (this density matrix may be understood
as a reduced density matrix by integrating over all other momenta).
Even though the many-body system is highly entangled in terms of wave
functions (Slater determinant), at zero
temperature this entanglement is hidden by the occupation number representation.
We may however, consider a subspace within each momentum value.

\begin{figure}[t]
\includegraphics[width=0.6\columnwidth]{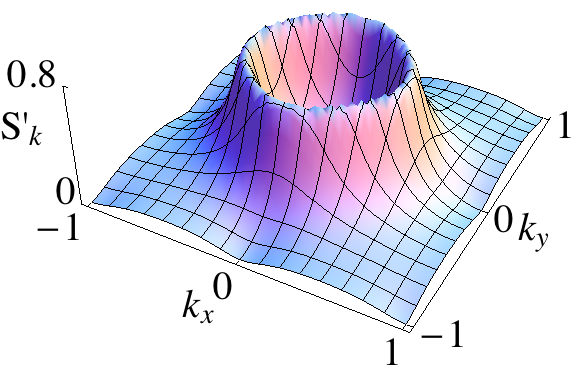}
\includegraphics[width=0.6\columnwidth]{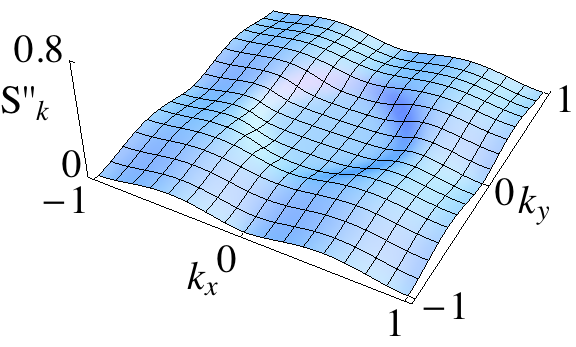}
\caption{\label{fig82}
(Color online) Entropy as a function of momentum for $M_z=4, \veps_F=2$ for the
reduced density matrices $\rho^{\prime}$ and $\rho^{\prime \prime}$, respectively.
}
\end{figure}

\begin{figure}[t]
\includegraphics[width=0.8\columnwidth]{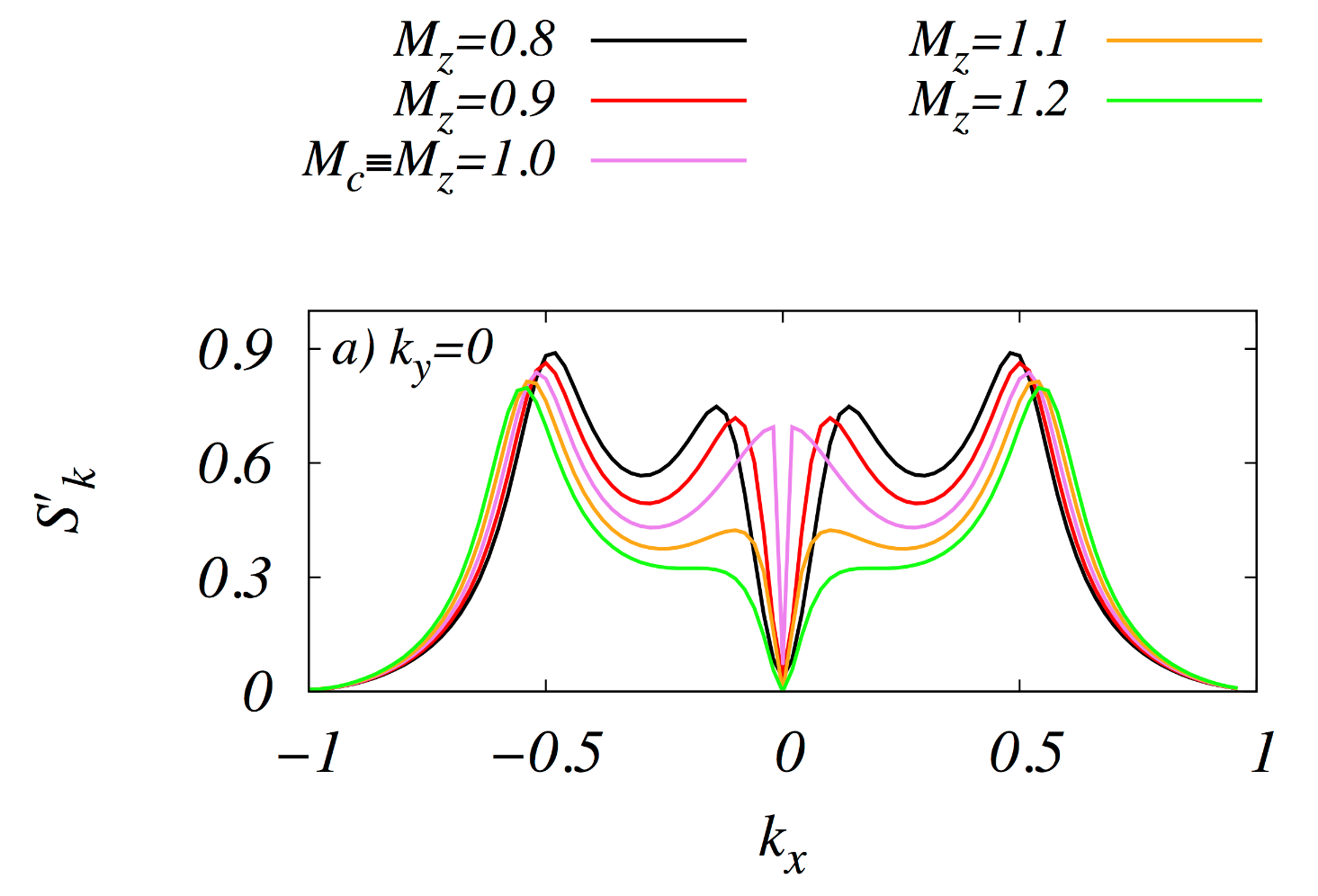}
\includegraphics[width=0.8\columnwidth]{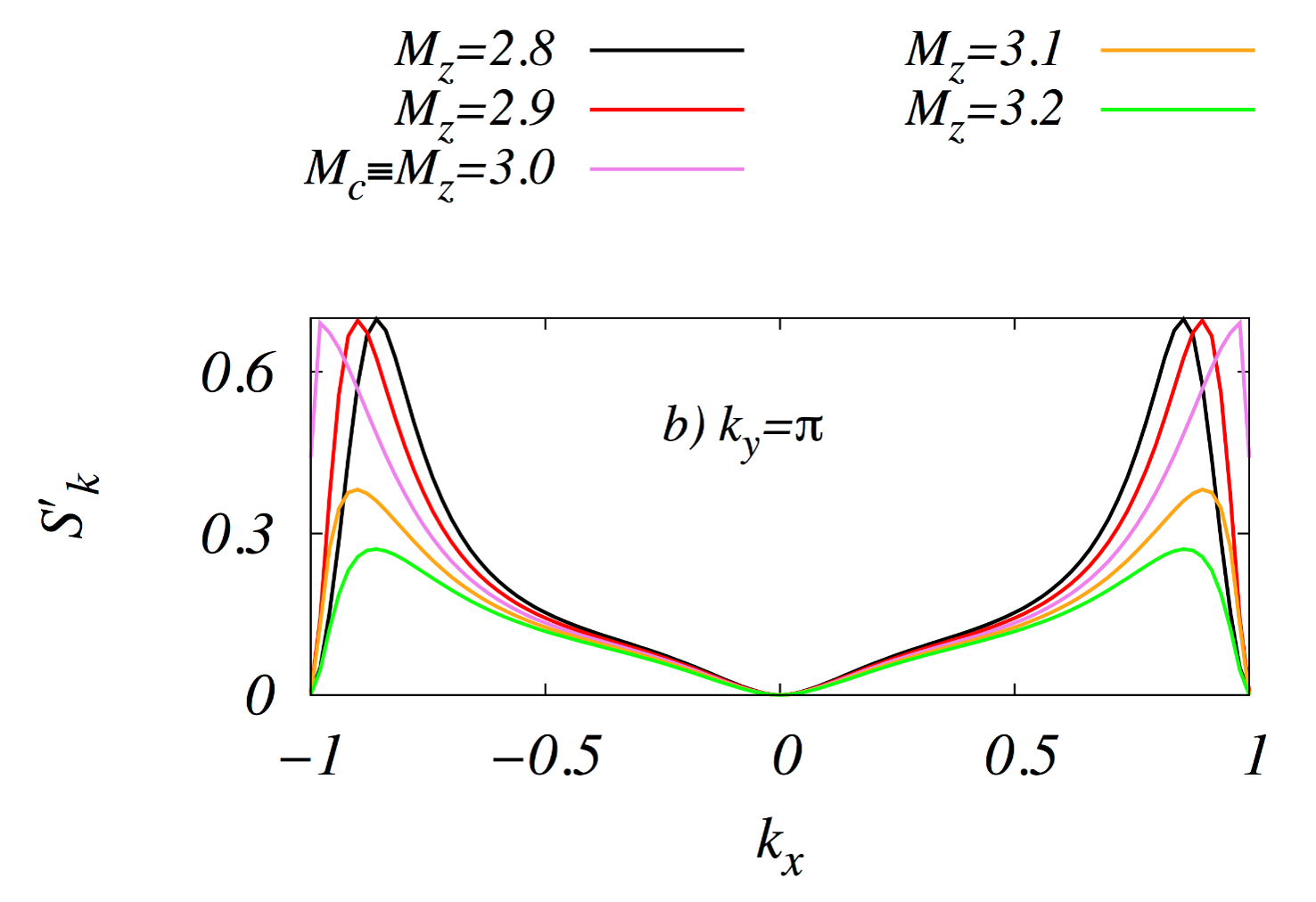}
\caption{\label{fig83}
(Color online) Entropy as a function of $k_x$ for a) $\veps_F=-3$ and $k_y=0$ and b) for
$\veps_F=1$ and $k_y=\pi$ for values of $M_z$ around the topological transitions.
}
\end{figure}

The basis states lead to a density matrix expressed as
\[
\rho(n_{k,\uparrow},n_{-k,\uparrow},n_{k,\downarrow},n_{-k,\downarrow};
n_{k,\uparrow}^{\prime},n_{-k,\uparrow}^{\prime},n_{k,\downarrow}^{\prime},n_{-k,\downarrow}^{\prime})
\]
where each number operator takes the values $0,1$.
The diagonalization of the Hamiltonian was obtained through a unitary transformation in the way
\be
\tilde{H}_d= {\bs Q} \tilde{H} {\bs Q}^{-1}
\ee
This transformation also diagonalizes the density matrix,
which allows us to calculate the diagonal form of it, $\rho_d$, through eq. (\ref{eq:eigval-DM}).
Therefore, having diagonalized the Hamiltonian we can obtain the density matrix
\be
\rho={\bs Q}^{-1} \rho_d {\bs Q}
\ee

We may now construct a $k$-subspace density matrix with dimension $4\times 4$ in
a reduced space by tracing out part of the degrees of freedom.
If we define as part A the $k$ momenta (with up and
down spin) and a part B with momenta $-k$ (with up and down spin), the reduced density matrix is
obtained tracing out the degrees of freedom, for instance, of part B, and we are left with a density
matrix for part A. This choice gives the entanglement between A and B which is the entanglement between
electrons with momentum $k$ and momentum $-k$, characteristic of superconductivity, including
the singlet and the triplet pairings.

Let us call $\rho^{\prime}$ the reduced density matrix of dimension $4\times 4$. It can be obtained
element by element as
\bea
& & \rho^{\prime}(n_{k,\uparrow},n_{k,\downarrow};
n_{k,\uparrow}^{\prime},n_{k,\downarrow}^{\prime}) =
\sum_{n_{-k,\uparrow},n_{-k,\downarrow}} \nonumber \\
& & \rho(n_{k,\uparrow},n_{-k,\uparrow},n_{k,\downarrow},n_{-k,\downarrow};
n_{k,\uparrow}^{\prime},n_{-k,\uparrow},n_{k,\downarrow}^{\prime},n_{-k,\downarrow}) \nonumber \\
& &
\eea
This density matrix gives information on the entanglement between electrons
of opposite momenta.

Alternatively, we may also define
\bea
& & \rho^{\prime \prime}(n_{k,\uparrow},n_{-k,\uparrow};
n_{k,\uparrow}^{\prime},n_{-k,\uparrow}^{\prime}) =
\sum_{n_{k,\downarrow},n_{-k,\downarrow}} \nonumber \\
& & \rho(n_{k,\uparrow},n_{-k,\uparrow},n_{k,\downarrow},n_{-k,\downarrow};
n_{k,\uparrow}^{\prime},n_{-k,\uparrow}^{\prime},n_{k,\downarrow},n_{-k,\downarrow}) \nonumber \\
& &
\eea
and now, this density matrix gives information on the entanglement between electrons
of opposite spins.

The diagonalization of these reduced density matrices leads to 4 eigenvalues. Typically the
two smaller eigenvalues take very small values and, since they are normalized to one,
the second largest eigenvalue is basically the difference to one of the largest eigenvalue.
As shown before \cite{us2} the largest eigenvalue takes values close to unity in most of the
Brillouin zone, except near the Fermi surface, where it decreases in value. This is shown in
Fig. \ref{fig81}: in the first panel we plot the lowest positive energy band obtained
from the solution of the Hamiltonian, eq. (\ref{bdgbands}) and in the second panel we
plot the largest eigenvalue of the reduced density matrix $\rho^{\prime}$. Further information
may be obtained calculating the entropy of this $k$-subspace reduced density matrix, defined as
\be
S_k^{\prime}=-\sum_{n=1}^4 \Lambda'_{k,n} \ln (\Lambda'_{k,n})
\ee
The results for the entropy are shown in the third panel of Fig. \ref{fig81} for the same
set of parameters, $M_z=2,\veps_F=0$. The decrease of the largest eigenvalue is matched
by the increase of the entropy, which therefore locates the region of the transition in
momentum space between the occupied states and the unoccupied states defining the Fermi surface.

In Fig. \ref{fig82} we compare the entropy between the reduced density matrices
$\rho^{\prime}$ and $\rho^{\prime \prime}$ at another point in the phase diagram,
$M_z=4, \veps_F=2$ ($C=-1$). The entropy is clearly larger in the case of the entanglement
between opposite momenta, as expected from a superconductor.

It is however, more interesting to study the entropy associated with the reduced density matrix
$\rho^{\prime}$ as one approaches a topological transition. In Fig. \ref{fig83} we consider
the dependence of the entropy with momentum for different values of the chemical potential,
$\veps_F=-3, \veps_F=1$, for different values of the magnetization, $M_z$, as one crosses
the transition lines between a $C=0$ phase and a $C=1$ phase and between a $C=-2$ and a
$C=-1$ phases. As with other signatures the time-reversal momentum values
play a special role associated with the transition lines. In the top panel of Fig. \ref{fig83}
there is a change of behavior around momentum $(0,0)$ and in the lower panel around momentum
$(\pi,\pi)$.
From the data analysis, we can say that
in the $C=0$ phase the entropy has a small, fixed but finite value,
at zero momentum.
As one crosses the transition to the $C=1$ phase the entropy vanishes at this momentum. Also,
as one approaches the transition line the entropy slope increases around zero momentum.
After the transition the slope decreases considerably and the entropy is smaller than in the
phase with zero Chern number. The lower panel shows a similar behavior but
around momentum $(\pi,\pi)$,
the slope becomes quite large at the transition and then decreases. Before the
transition, as the magnetization is increasing, the entropy is also finite (but rather small)
and after the transition it vanishes at that momentum. Also, the magnitude of the entropy
is smaller after the transition, since the magnetic field is larger.

In Appendix B we present results for the eigenvector of the largest eigenvalue of the
$k$-subspace reduced density matrix.

\begin{figure}[h]
\includegraphics[width=0.9\columnwidth]{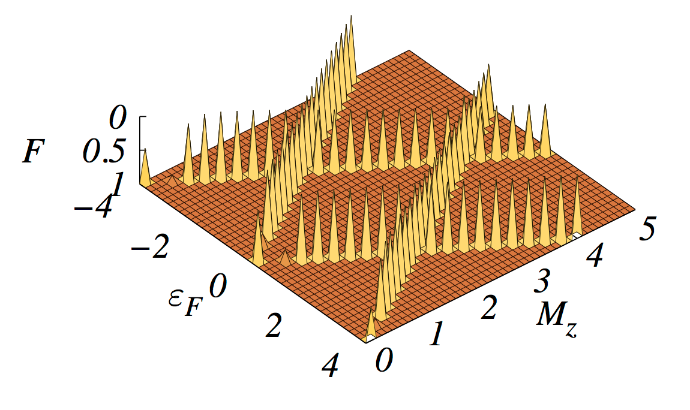}
\caption{\label{fig8}
(Color online) Total fidelity as a function of chemical potential and magnetization.
The fidelity was calculated using two points from parameter space, $M_z$ and $M_z + \delta M_z$, with $\delta M_z = 0.01$.
Note that the vertical scale is inverted.
}
\end{figure}


\section{Fidelity and fidelity spectrum}
\label{sec:Fidelity}

The quantum
fidelity between two pure states (for two sets of parameters) is
the absolute value of the overlap between the ground states for the two sets of parameters.
In general, the quantum fidelity \cite{Jozsa} between two states characterized by two density
matrices $\rho_1$ and $\rho_2$, may be defined as the trace of the fidelity operator,
${\cal F}$, as
$F(\rho_1,\rho_2) = \mbox{Tr} {\cal F}= \mbox{Tr} \sqrt{ \sqrt{\rho_1} \rho_2 \sqrt{\rho_1}}$.

The fidelity operator $\mathcal{F}$ can be studied using different basis states, associated with
different representations, such as position, momentum, energy or charge and spin.
Rewriting the fidelity operator in these different representations, allows us to
look more directly at the specific
relevant modes that
participate more actively in the
critical phenomena accompanying the phase transition. In this way, we can obtain a more complete and physical
characterization of the phase transition and of its underlying physics
mechanisms.
While the entanglement spectrum has some relation to the energy spectrum of the edge states
or even bulk states, the fidelity spectrum contains information about which
eigenvalues have a larger contribution to the distinguishability between quantum states \cite{us2}.

Considering two density matrices, that result from the momentum space partition,
for two points in parameter space, we can write that
\begin{equation}
\label{ }
	{\rho_1}_{k} {\bm Q_1}_{k} = {\bm Q_1}_{k} {\bm \Lambda_1}_{k} \quad ; \quad
        {\rho_2}_{k} {\bm Q_2}_{k} = {\bm Q_2}_{k} {\bm \Lambda_2}_{k}
\end{equation}

As mentioned above, the fidelity between two states, characterized by two density
matrices ${\rho_1}_{k}$ and ${\rho_2}_{k}$, may then be defined as the trace of the fidelity operator,
${\cal F}_{k}$,
\be
F_{k}({\rho_1}_{k},{\rho_2}_{k}) = \mbox{Tr} {\cal F}_{k}= \mbox{Tr} \sqrt{ \sqrt{ {\rho_1}_{k} } {\rho_2}_{k} \sqrt{ {\rho_1}_{k} } }
\label{eq1}
\ee
The square root of the density matrix can be written as
\begin{equation}
\label{ }
	\sqrt{{\rho_1}_{k}} = {\bm Q_1}_{k} \sqrt{ {\bm \Lambda_1}_{k}} {\bm Q_1}_{k}^{-1}
\end{equation}
and
\begin{equation}
\label{ }
	{{\cal F}_{k}}^{2} = \sqrt{{\rho_1}_{k}}  {\rho_2}_{k}  \sqrt{{\rho_2}_{k}}
\end{equation}
Diagonalizing
\begin{equation}
\label{ }
	{\cal F}_{k} = \bm{U}_{k}  \bm{f}_{k}  \bm{U}_{k}^{-1}
\end{equation}
we write that
\begin{equation}
\label{ }
	{\cal F}_{k,n}	= \bm{U}_{k,n}  f_{k,n}  \bm{U}_{k,n}^{-1} \quad ; \quad n = 1, \ldots, 16
\end{equation}
Therefore the $k$-fidelity is obtained as
\begin{equation}
\label{ }
	F_{k}	= \mathrm{Tr} ( {\cal F}_{k} )
		= \sum\limits_{n} f_{k,n}
\end{equation}
and the total fidelity is finally obtained as
\begin{equation}
\label{ }
	F = \prod\limits_{k} F_{k}
\end{equation}

We begin by considering that the two density matrices,
${\rho_1}_{k}, {\rho_2}_{k}$, are for two points nearby in parameter space.
As the fidelity measures the distinguishability between different states, as one
approaches a transition one of the density matrices is calculated on one side of
transition and the other is either at the transition point or in the other phase
(depending on the step in parameter space) and the fidelity shows a minimum.

\begin{figure}[h]
\includegraphics[width=0.8\columnwidth]{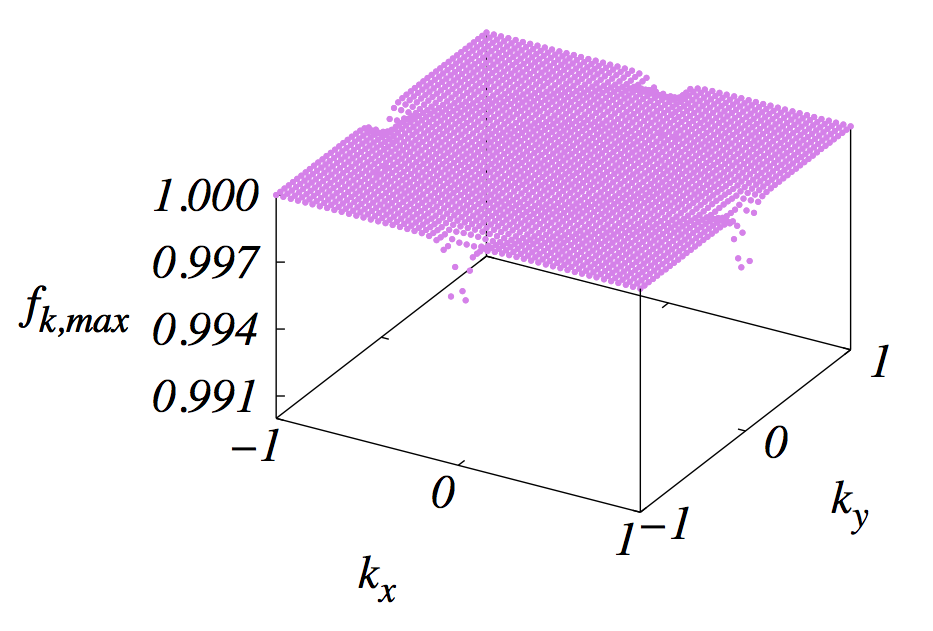}
\caption{\label{fig9}
(Color online) Highest eigenvalue of the $k$-fidelity spectrum at a transition point $\veps_F=-1,M_z=1$
as a function of momentum.
The fidelity was calculated using two points from parameter space, $\veps_F$ and $\veps_F + \delta \veps_F$,
with $\delta \veps_F = 0.01$.
The transition occurs for momentum $\boldsymbol{k}=(\pi,0)$ and equivalent
points, as shown by the decrease of the fidelity in the neighborhood of the singular momenta values.
}
\end{figure}

In Fig. \ref{fig8} we present results for the total fidelity as a function of the chemical
potential and the magnetization. Far from the transition points the fidelity is close to one,
as the two states described by the two density matrices are very similar. The minima in the
fidelity faithfully track the transitions previously described (note that the vertical axis
is upside down). As for the entropy, fixing the magnetization at $M_z=0.1$ the $k$-fidelity
faithfully singles out the singular momenta values.

The transitions may also be signaled by the decrease of the highest eigenvalue of the
$k$-fidelity spectrum at the transition point, when the two density matrices are calculated
at points in parameter space surrounding the transition. This is shown in Fig. \ref{fig9} for
a particular example, $\veps_F=-1,M_z=1$, for which the transition line is associated with the
singular points $\boldsymbol{k}=(\pi,0)$, and equivalent points in the Brillouin zone.

Complementary, interesting information, can be obtained taking the two density matrices in the
fidelity expression, corresponding to points in parameter space that are far apart. This allows
to compare different phases, and not specifically to detect the locations of the phase transitions.
It provides interesting information about the nature and momenta values responsible for the
transitions.
Also, it describes the distinguishability between two phases in momentum space.

\begin{figure}[h]
\includegraphics[width=0.7\columnwidth]{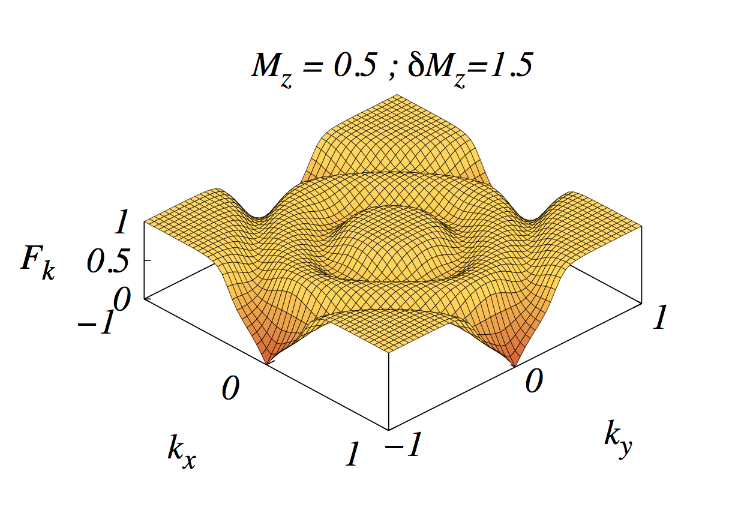}
\includegraphics[width=0.7\columnwidth]{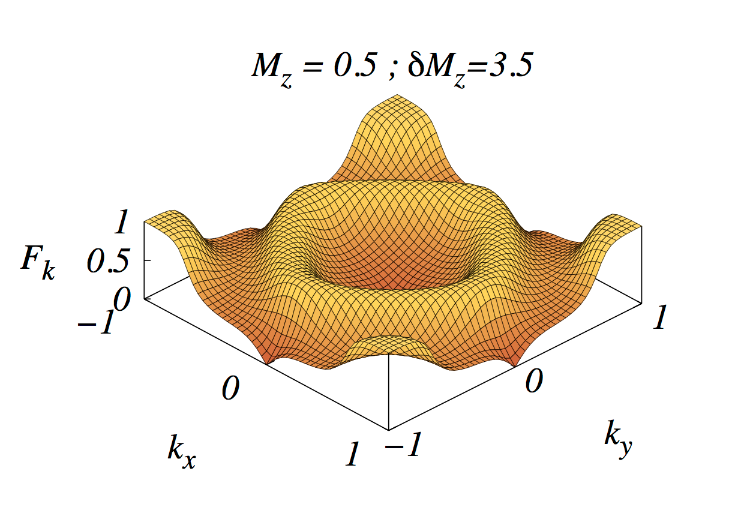}
\includegraphics[width=0.7\columnwidth]{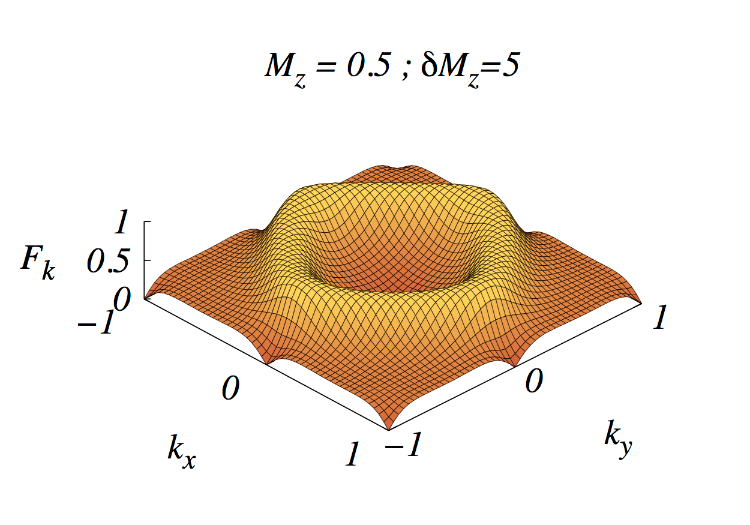}
\caption{\label{fig10}
(Color online) $k$-fidelity operator spectrum $F_k$ for $\rho_A$ corresponding to $\veps_F=-1, M_z=0.5$ where
the Chern $C=0$,
and $\rho_B$ corresponding to $\veps_F=-1$ as well and $M_z=2,C=-2$, $M_z=4,C=-1$ and $M_z=5.5,C=0$, respectively.
}
\end{figure}

\begin{figure}[h]
\includegraphics[width=0.7\columnwidth]{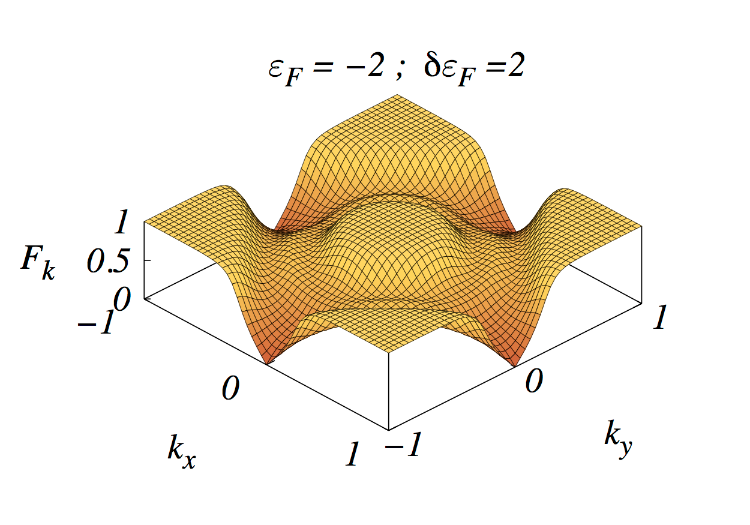}
\includegraphics[width=0.7\columnwidth]{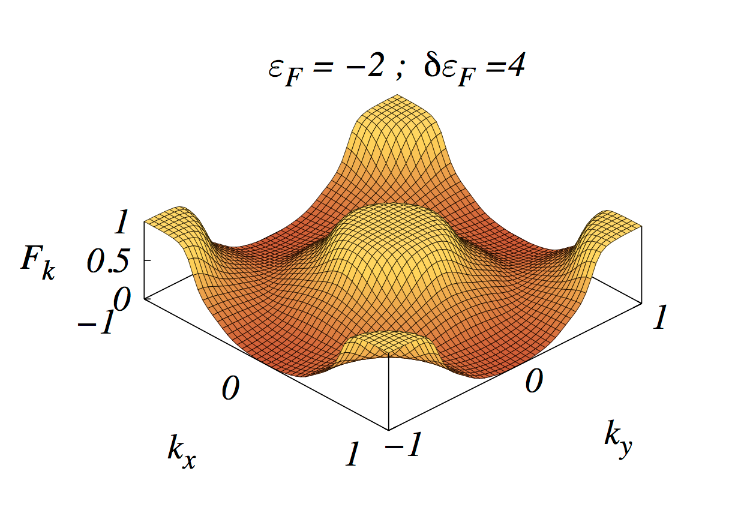}
\includegraphics[width=0.7\columnwidth]{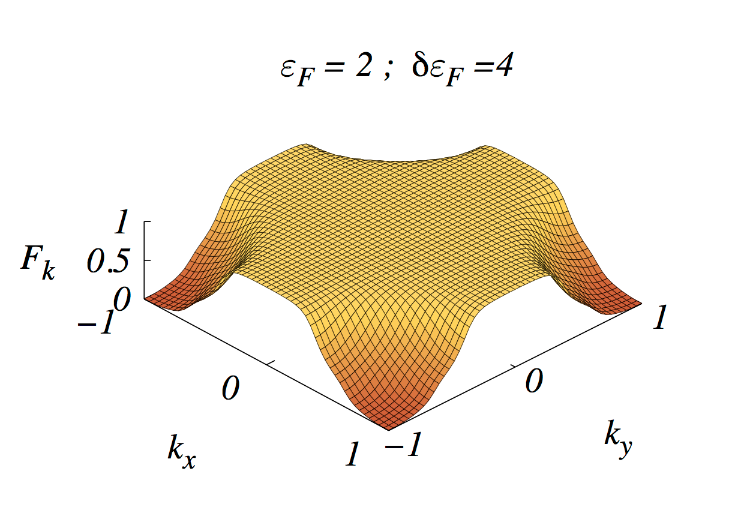}
\caption{\label{fig11}
(Color online) $k$-fidelity operator spectrum $F_k$ for $\rho_A$ corresponding to $\veps_F=-2, M_z=0.5$ where
the Chern $C=0$,
and $\rho_B$ corresponding to $M_z=0.5$ as well and $\veps_F=0,C=-2$, $\veps_F=2,C=0$ and
at bottom panel $\rho_A$ corresponding to $\veps_F=2, M_z=0.5$ where
the Chern $C=0$ and $\rho_B$ corresponding to $M_z=0.5,\veps_F=6,C=0$, respectively.
}
\end{figure}

In Figs. \ref{fig10},\ref{fig11} we present results for the $k$-fidelity operator spectrum,
where the two density matrices define two states in parameter space that are deep inside different
topological phases. In Fig. \ref{fig10} we consider one density matrix at a state specified by
a point in a trivial phase with $C=0$ (specifically $\veps_F=-1,M_z=0.5$) and the other
density matrix corresponds to states at various magnetizations, along the vertical dashed line
of Fig. \ref{fig1}, for which the Chern numbers are $C=-2,C=-1,C=0$ and the magnetization is
$M_z=2,M_z=4,M_z=5.5$, respectively. As we move along that line in parameter space, we cross,
in sequential order, lines characterized by momenta $\boldsymbol{k}=(\pi,0),\boldsymbol{k}=(0,0),
\boldsymbol{k}=(\pi,\pi)$,
respectively.
The $k$-space fidelity highlights the differences between the various phases.
Its deviations from unity throughout the Brillouin zone are significant, even when
the phases have the same Chern number, as a consequence of the different band-fillings.
Singular transition points translate to zeros in the $k$-fidelity operator spectrum.
As the top panel shows, the $k$-fidelity operator spectrum vanishes at the momenta $\boldsymbol{k}=(\pi,0)$
and equivalent points (two of them are independent). The middle panel shows in addition a zero
at the center of the Brillouin zone $\boldsymbol{k}=(0,0)$. Consequently the last panel shows zeros at
all singular momentum locations. The $k$-fidelity operator therefore signals the minimum number of transitions
that must occur when going from one trivial phase to the other trivial phase.

Similar results are shown in Fig. \ref{fig11}. In this case we also start from the same trivial
phase with $C=0$ ($\veps_F=-2,M_z=0.5$) and consider a sequence of density matrices, associated
with states along the horizontal dashed line of Fig. \ref{fig1}. In this instance we consider
in the top panel that the second density matrix corresponds to $M_z=0.5,\veps_F=0$, $C=-2$. A line
associated with $\boldsymbol{k}=(\pi,0)$ is crossed. In the middle panel we consider that the second density
matrix is at point $M_z=0.5,\veps_F=2$, $C=0$. A second transition line at the same momentum is now crossed,
which implies a zero of quadratic order, since the same momentum becomes gapless twice.
A similar result occurs if we consider a density matrix at $M_z=0.5,\veps_F=2$ and the other at the same
magnetization and $\veps_F=6$. In this case the momentum $\boldsymbol{k}=(\pi,\pi)$ contributes twice with
a gapless spectrum and the dispersion is now quadratic around that momentum (the corners of the Brillouin zone),
as shown in the bottom panel of this figure.

\section{Conclusions}
\label{sec:conclusion}

Various quantum information techniques have been used in this work to detect and distinguish topological phase
transitions in superconductors. In particular, the entanglement von Neumann entropy,
entanglement spectrum, fidelity and the recently introduced fidelity spectrum.
In addition to a faithful identification of the transitions, these methods clearly
signal the relevant modes responsible for the transitions.
The analysis of the highest weight eigenvector components clarifies the nature of
each phase and the contents associated with a transition between two phases. Also, allows to
distinguish between a trivial and a topological phase and between the $\mathbb{Z}_2$ and $\mathbb{Z}$ topological
phases. Specifically, for the p-wave superconductor
considered here, the spin triplet basis state has a particular role in the non-trivial
topological phases reaching saturation. A detailed analysis of the momentum dependence of the coefficients
of the basis states of the highest state eigenvector was carried out.
The analysis of the $k$-fidelity
operator spectrum allows a characterization of the sequence of the minimal number of
transitions that have to occur between two arbitrary points in the phase diagram.

\begin{figure*}[t]
\includegraphics[width=0.35\textwidth]{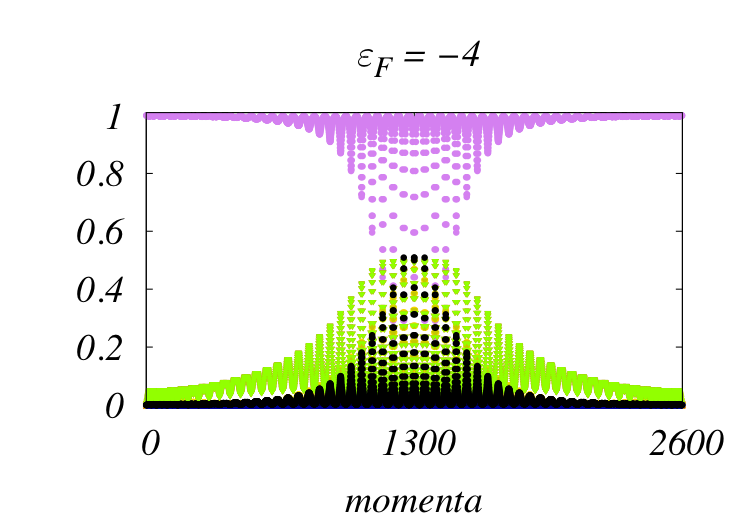}
\includegraphics[width=0.35\textwidth]{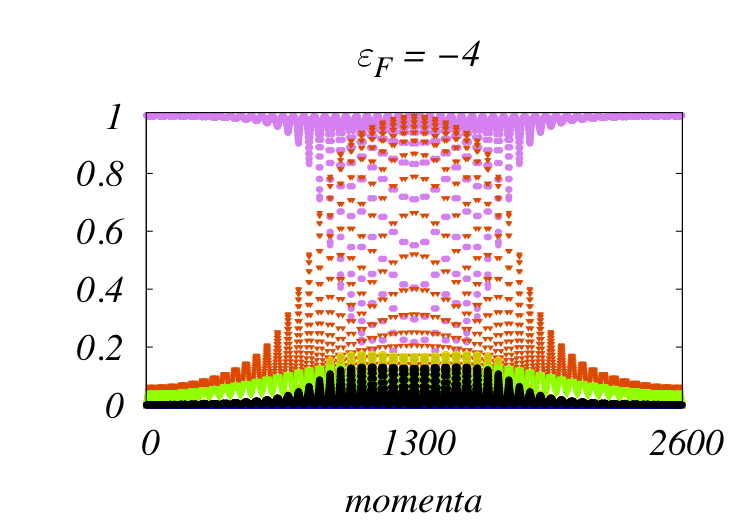}
\includegraphics[width=0.35\textwidth]{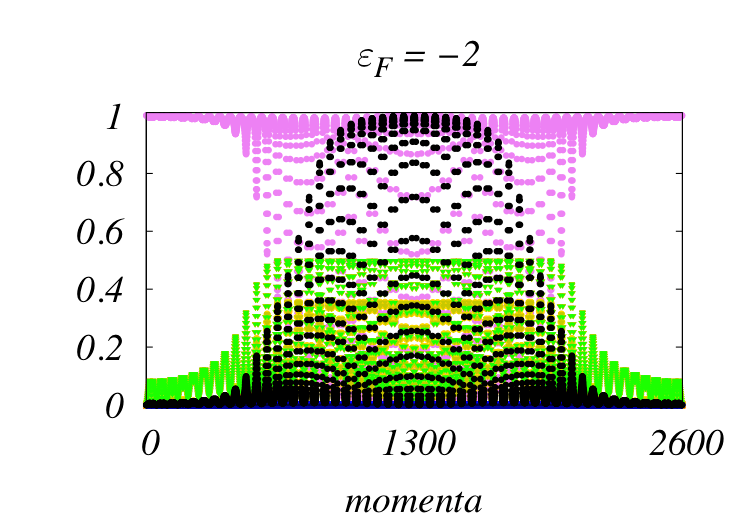}
\includegraphics[width=0.35\textwidth]{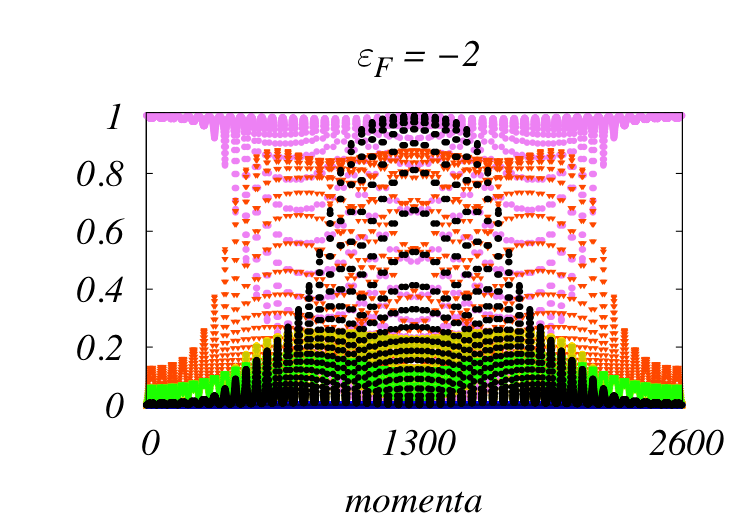}
\includegraphics[width=0.35\textwidth]{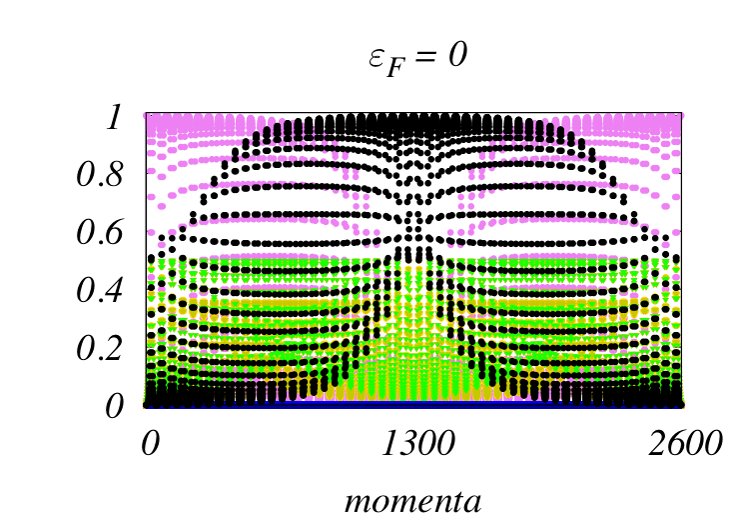}
\includegraphics[width=0.35\textwidth]{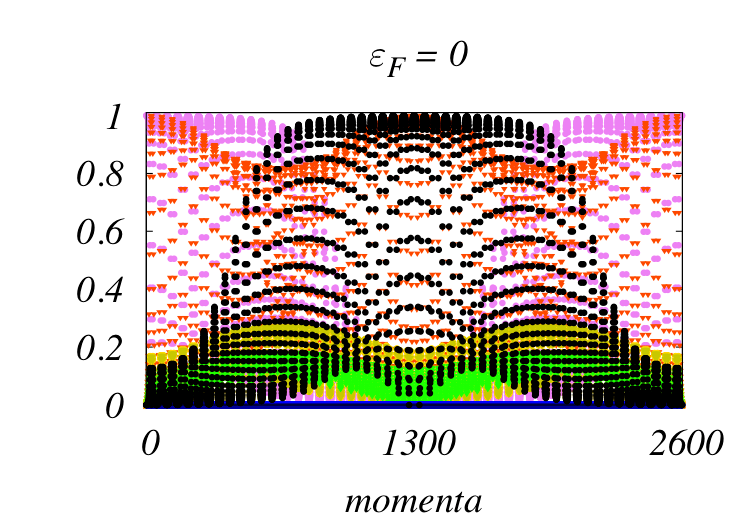}
\includegraphics[width=0.35\textwidth]{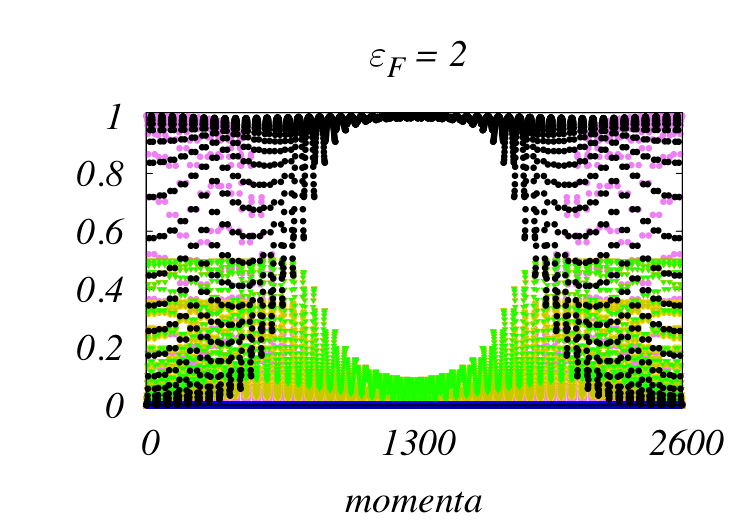}
\includegraphics[width=0.35\textwidth]{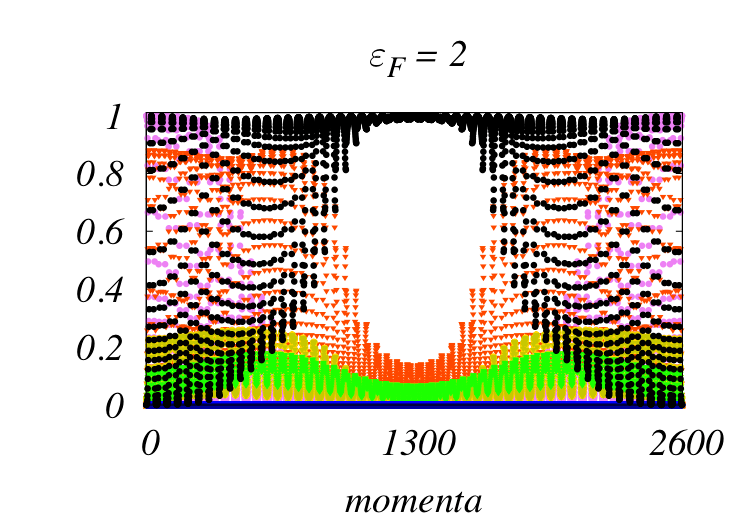}
\caption{\label{fig13}
(Color online) Absolute value of the coefficients of the basis states of the highest weight eigenvector as a function
of momenta values. In the left panels $M_z=0$ and in the right panels $M_z=1$. The rows are for the values
of $\veps_F=-4,-2,0,2$, respectively, from top to bottom.
The color codes of the coefficients
is the same as in Fig. \ref{fig7}.
}
\end{figure*}

\section*{Acknowledgements}

We thank discussions with V\'{\i}tor Rocha Vieira, Pedro Ribeiro, Miguel Ara\'ujo and Eduardo Castro and partial
support by the Portuguese FCT under grant PEST-OE/FIS/UI0091/2011.
T.P.O. would like to thank the Brazilian agency Coordena\c{c}\~ao de Aperfei\c{c}oamento
de Pessoal de N\'{\i}vel Superior (CAPES) the financial aid for study abroad.
P.D.S. acknowledges the hospitality and partial support from the Beijing Computational Science
Research Center where part of this work was carried out.

\appendix

\section{Highest weight density matrix eigenvector}

\begin{figure*}[t]
\includegraphics[width=0.25\textwidth]{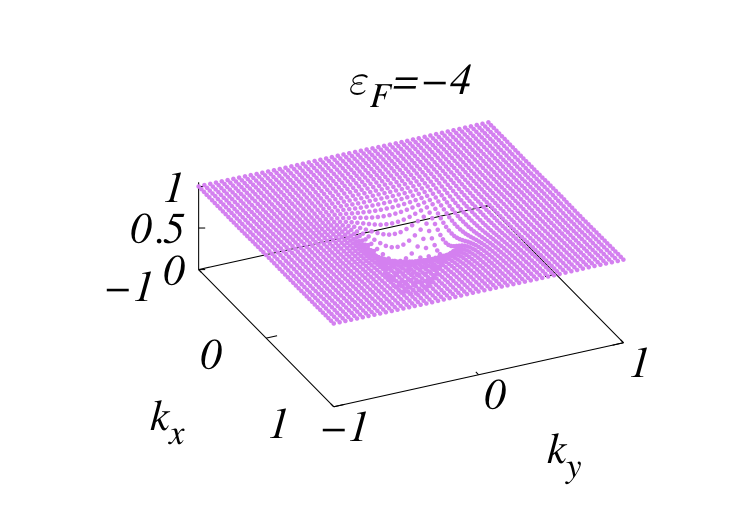}
\includegraphics[width=0.25\textwidth]{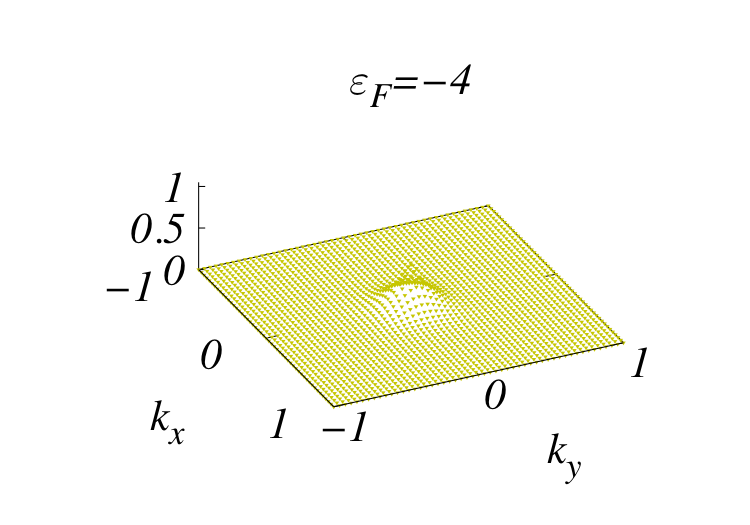}
\includegraphics[width=0.25\textwidth]{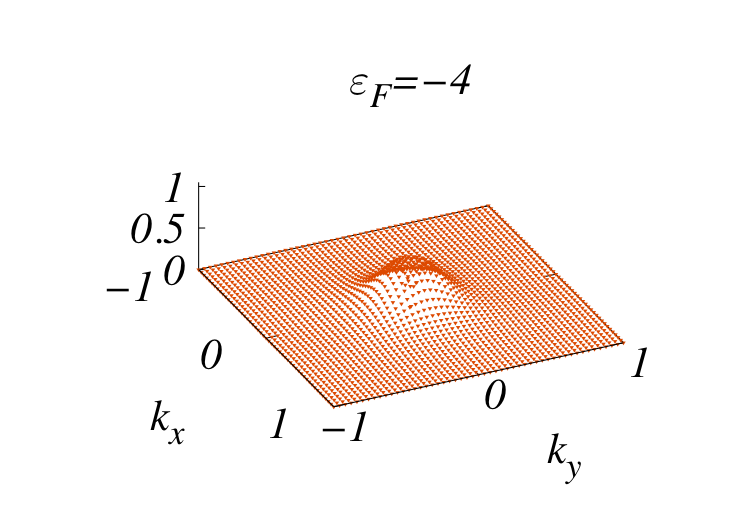}
\includegraphics[width=0.25\textwidth]{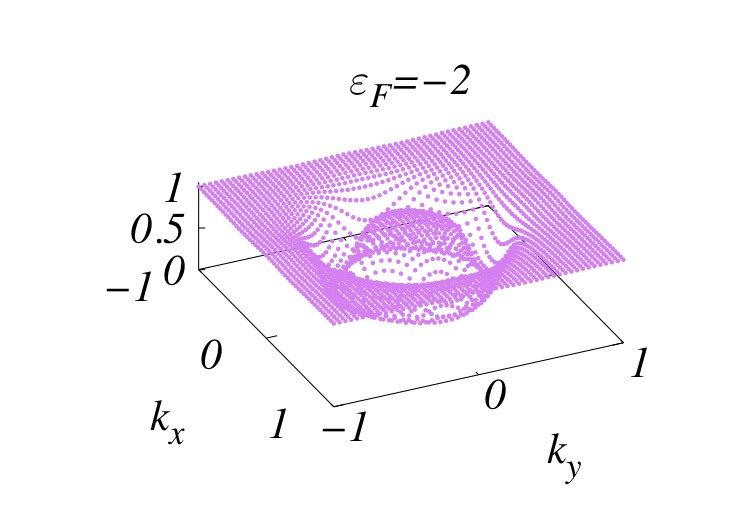}
\includegraphics[width=0.25\textwidth]{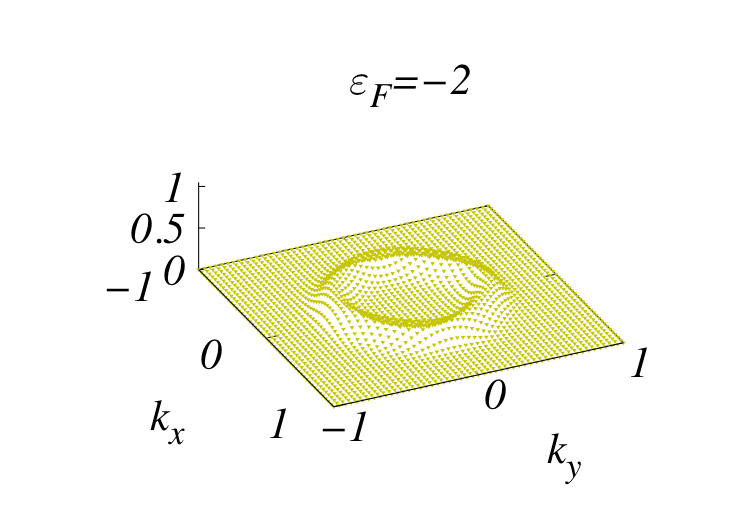}
\includegraphics[width=0.25\textwidth]{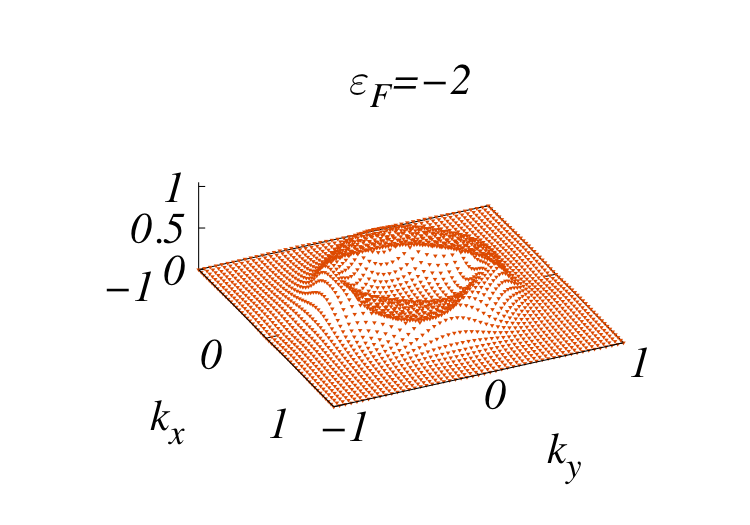}
\includegraphics[width=0.25\textwidth]{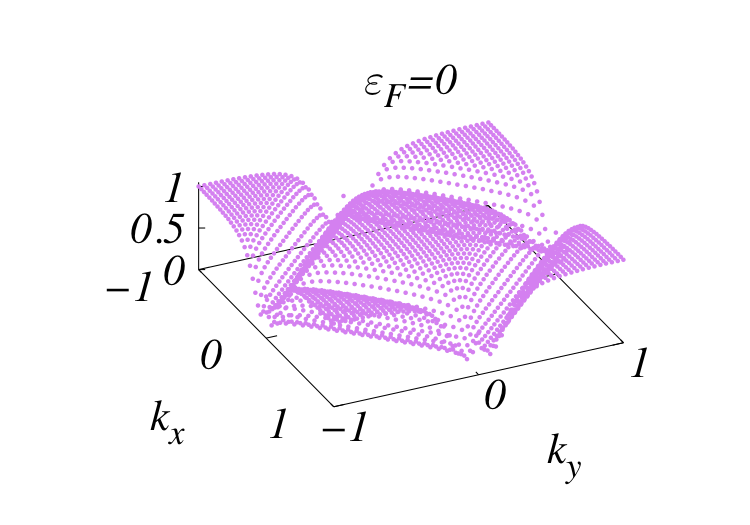}
\includegraphics[width=0.25\textwidth]{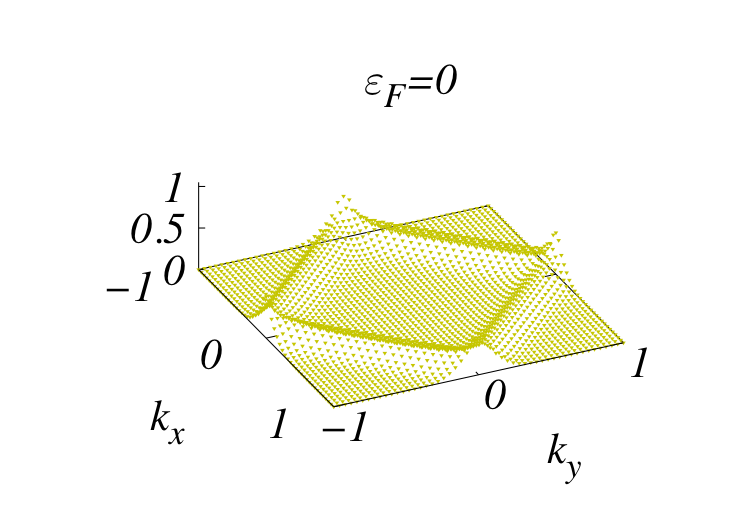}
\includegraphics[width=0.25\textwidth]{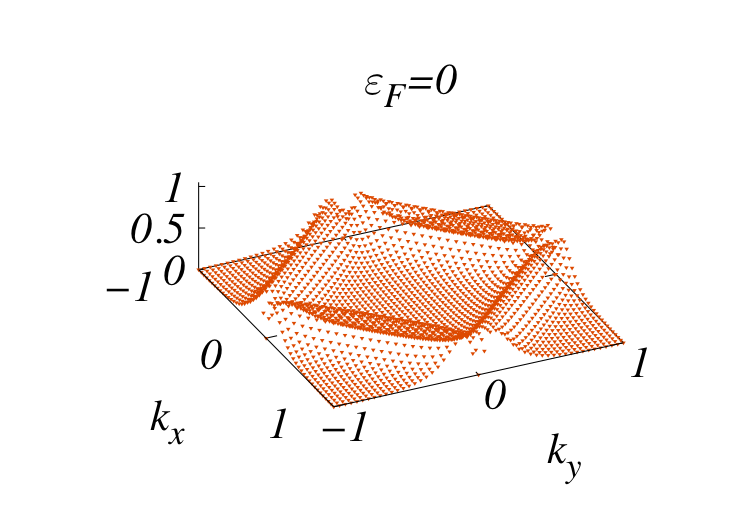}
\includegraphics[width=0.25\textwidth]{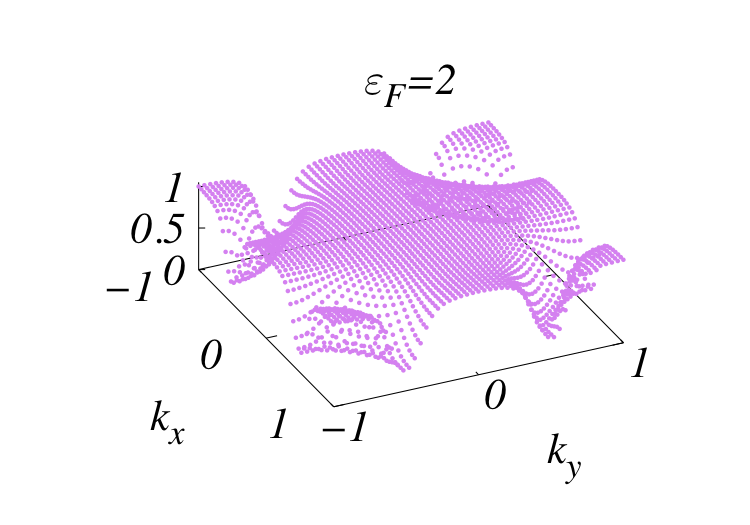}
\includegraphics[width=0.25\textwidth]{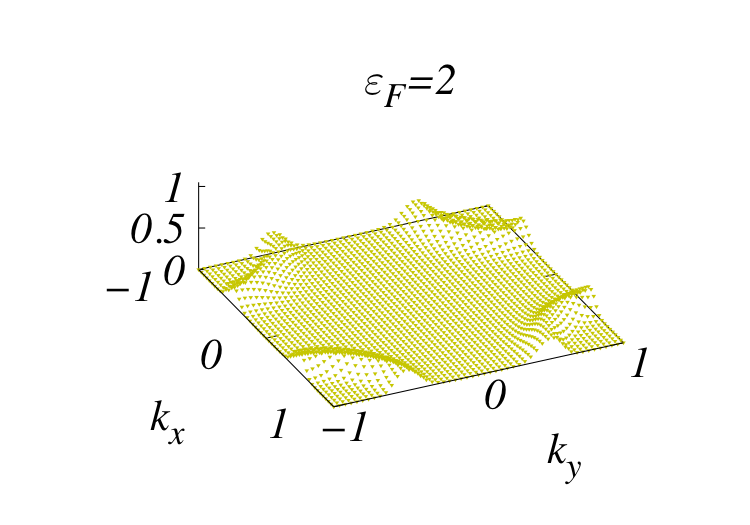}
\includegraphics[width=0.25\textwidth]{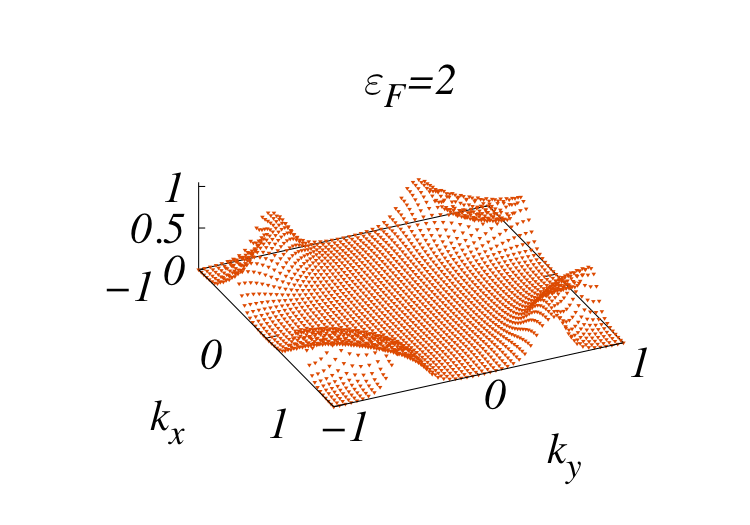}
\includegraphics[width=0.25\textwidth]{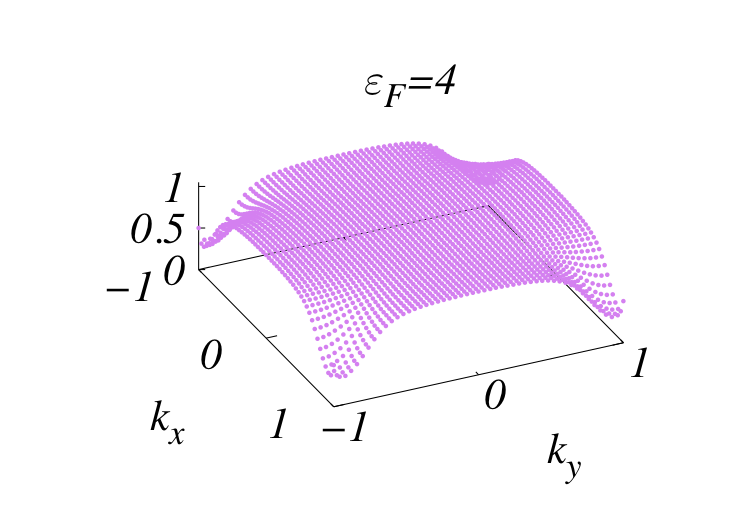}
\includegraphics[width=0.25\textwidth]{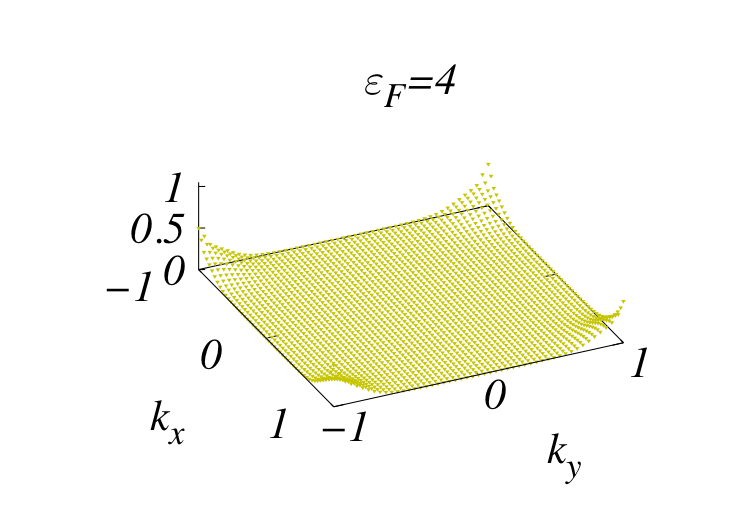}
\includegraphics[width=0.25\textwidth]{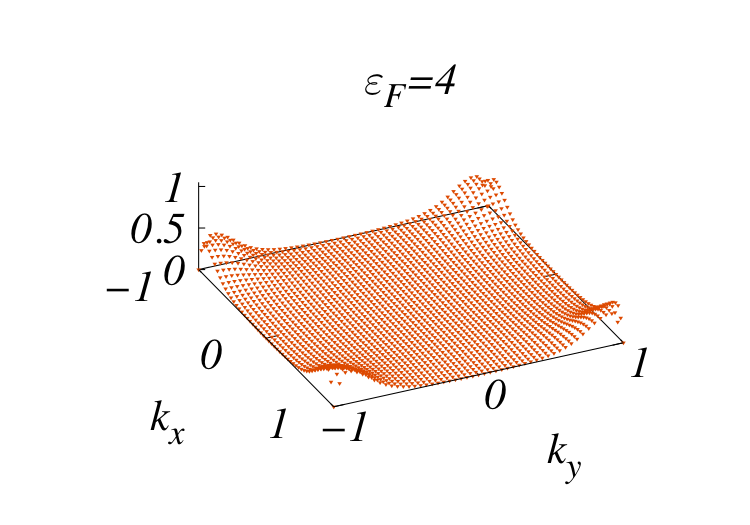}
\caption{\label{fig14}
(Color online) Sum of the absolute square values of the coefficients for the empty state and doubly occupied state (left panels),
for the spin singlet states (middle panels) and triplet states (right panels), for $M_z=0$ and for
$\veps_F=-4,-2,0,2,4$ from top to bottom,
 as a function of momentum.
}
\end{figure*}

In this appendix we study in detail the coefficients of the highest weight eigenvector.
We consider a sequence of states as a function of the chemical potential, $\veps_F$,
for different values of the magnetization $M_z$. We compare the results for $M_z=0$ and
$M_z=1$ for $\veps_F=-4,-2,0,2,4$, typically.
Along these sets of values the Chern number changes from topological phases
with $C=0$ to phases with finite Chern number. As discussed in the main text, even
though some phases have no Chern number, there are edge states except in the $C=0$ phases
with small magnetization and $|\veps_F|>4$, and large magnetization around $\veps_F=0$,
indicated in Fig. \ref{fig1}.
For instance, considering $M_z=1$ and changing the chemical potential, the Chern goes through a sequence
of values, $C=1,C=0,C=-2,C=0,C=1$, from non-trivial to trivial phases, crossing in sequence lines that
are singular at the momentum values $\boldsymbol{k}=(0,0),\boldsymbol{k}=(\pi,0), \boldsymbol{k}=(\pi,\pi)$.

In Fig. \ref{fig13} we present, in a rather compact way, the absolute values of the coefficients
of the 16 basis states, for the highest weight eigenvector, as a function of momentum, ordered sequentially
in the two-dimensional plane. We compare the results for $M_z=0$ and $M_z=1$ for different chemical potentials.
The color codes of the coefficients is the same as in Fig. \ref{fig7}.
Clear differences can be seen between the $\mathbb{Z}_2$ phases and the $\mathbb{Z}$ phases. For $\veps_F=-4$ and $M_z=1$
the system is in a phase with $C=1$ but for $M_z=0$ the system is at the frontier between a completely trivial
phase with $C=0$ and no edge states and a $\mathbb{Z}_2$ phase also with $C=0$ but edge states.
The dominant state is the state $|0000\rangle$ in most of momentum space. For $M_z=0$ the two spin orientations
of the triplet state, $|1100\rangle$ and $|0011\rangle$, are degenerate and are superimposed. The other state that
gives a significant contribution is the fully occupied state $|1111\rangle$.

\begin{figure*}[t]
\includegraphics[width=0.25\textwidth]{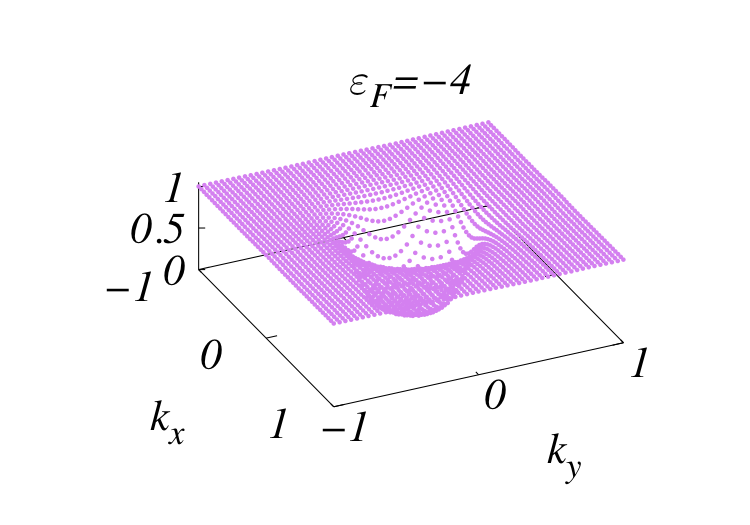}
\includegraphics[width=0.25\textwidth]{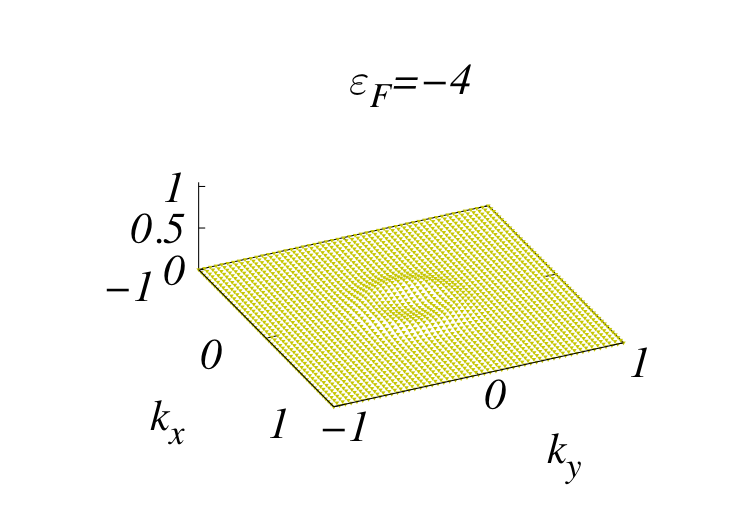}
\includegraphics[width=0.25\textwidth]{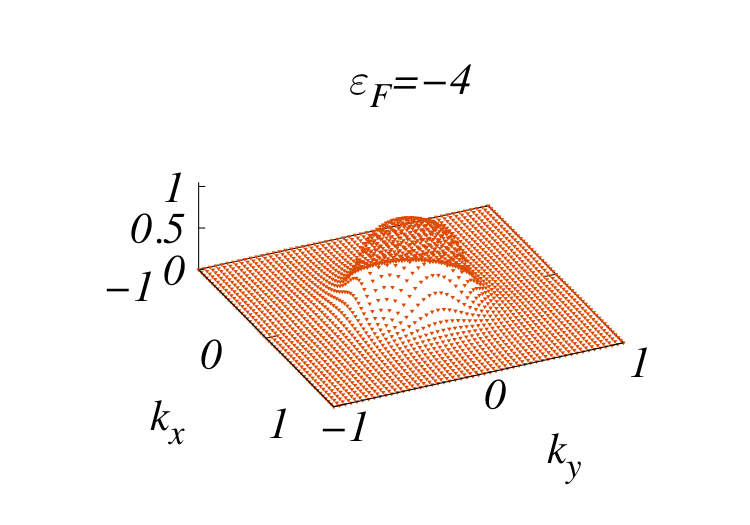}
\includegraphics[width=0.25\textwidth]{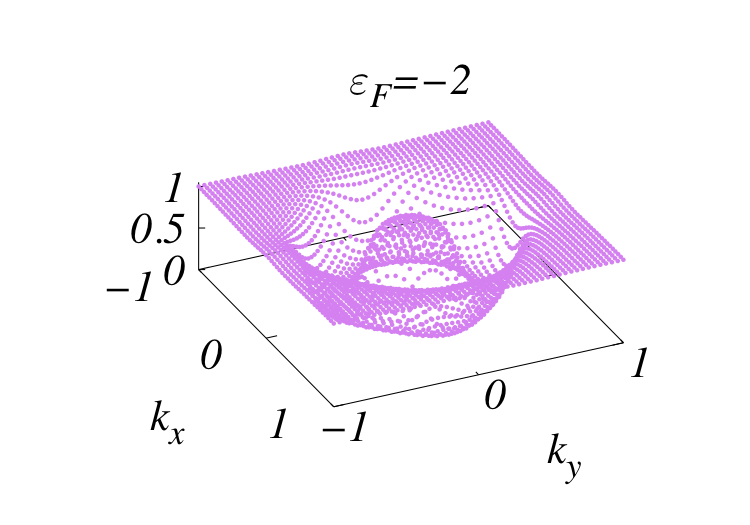}
\includegraphics[width=0.25\textwidth]{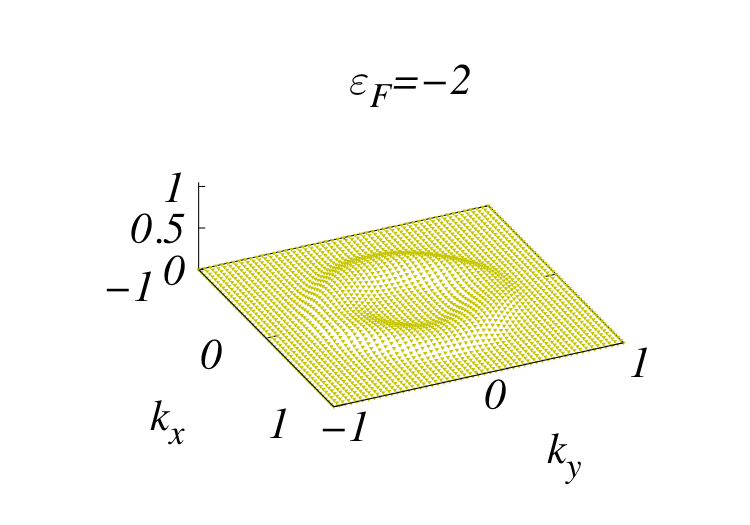}
\includegraphics[width=0.25\textwidth]{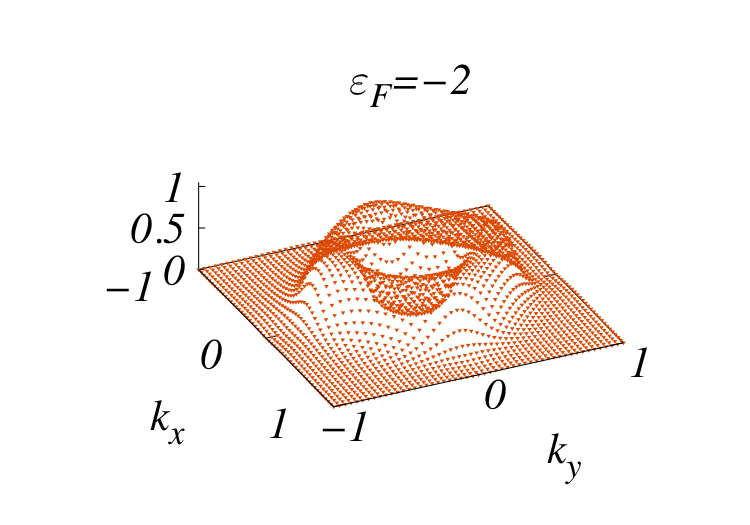}
\includegraphics[width=0.25\textwidth]{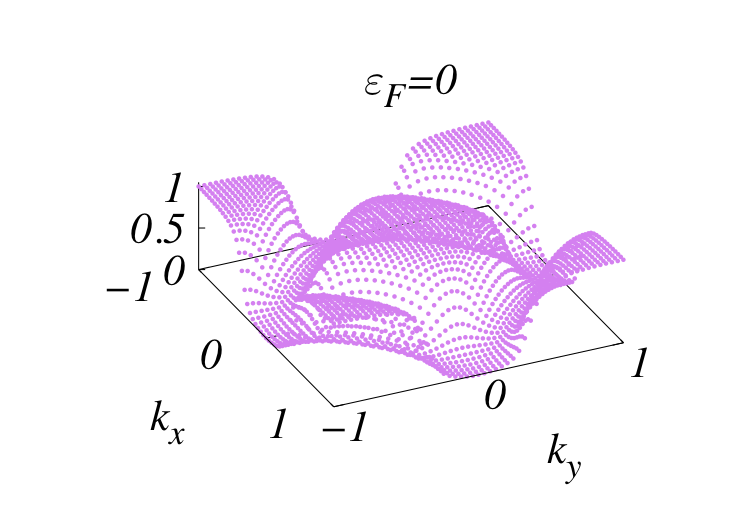}
\includegraphics[width=0.25\textwidth]{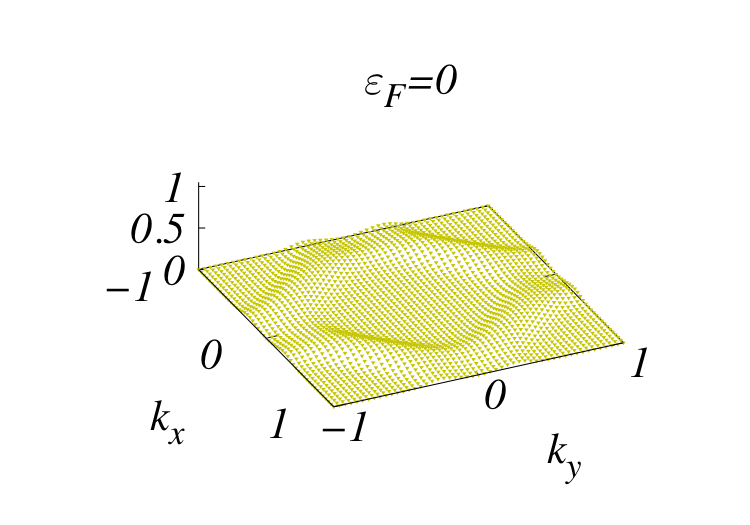}
\includegraphics[width=0.25\textwidth]{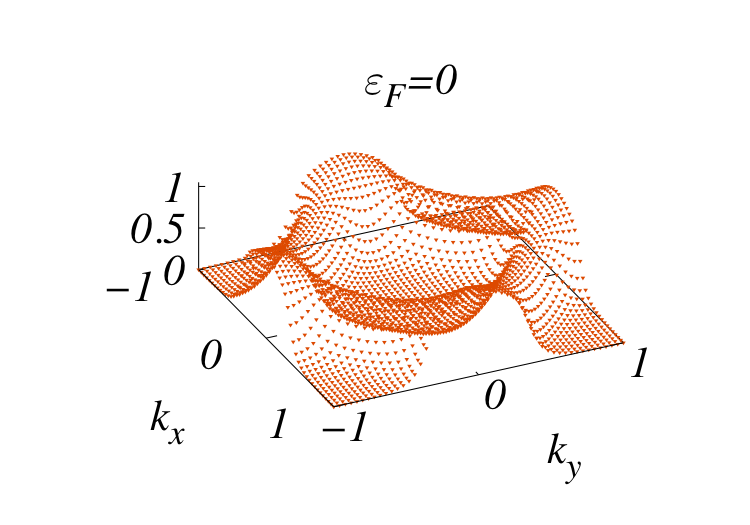}
\includegraphics[width=0.25\textwidth]{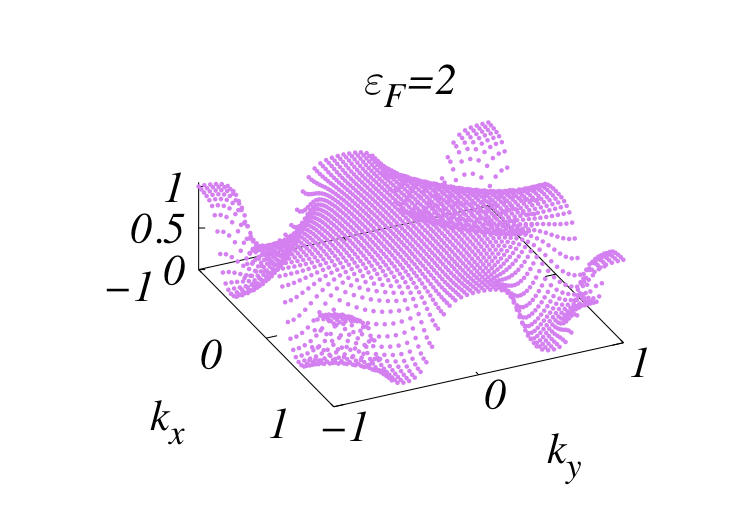}
\includegraphics[width=0.25\textwidth]{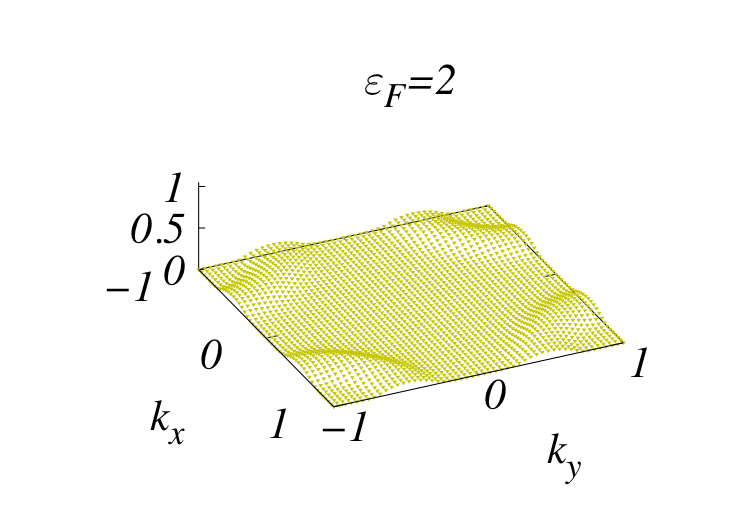}
\includegraphics[width=0.25\textwidth]{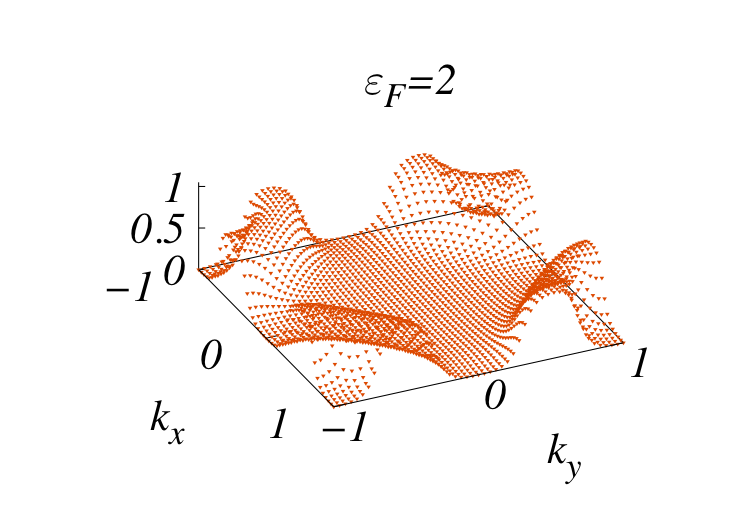}
\includegraphics[width=0.25\textwidth]{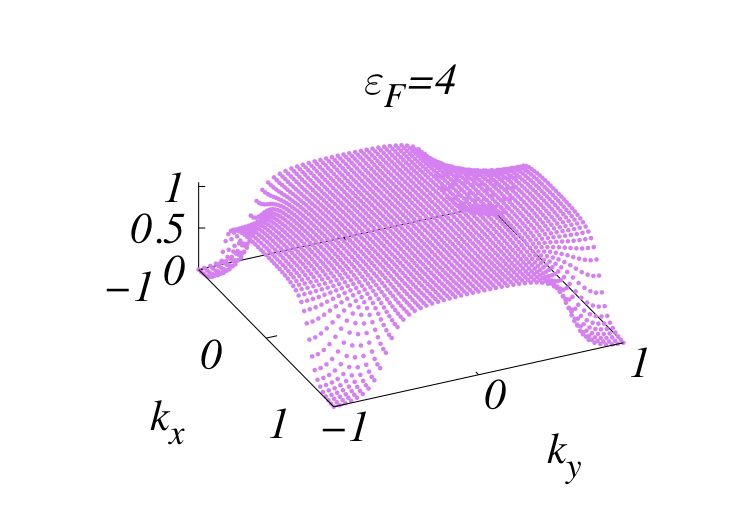}
\includegraphics[width=0.25\textwidth]{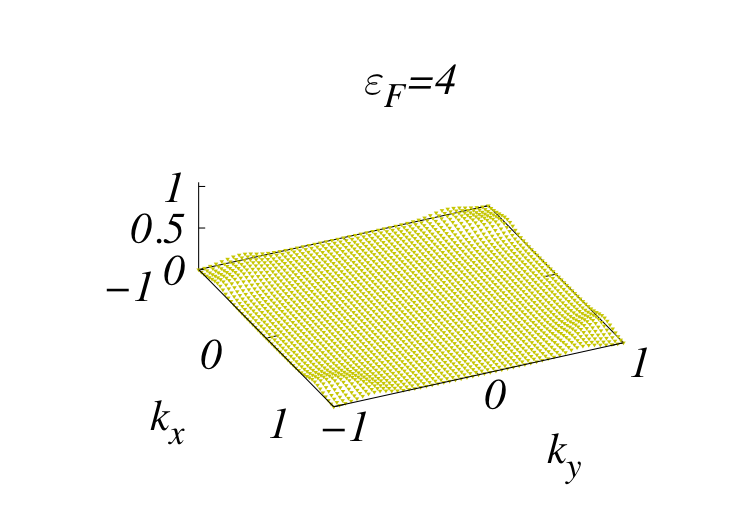}
\includegraphics[width=0.25\textwidth]{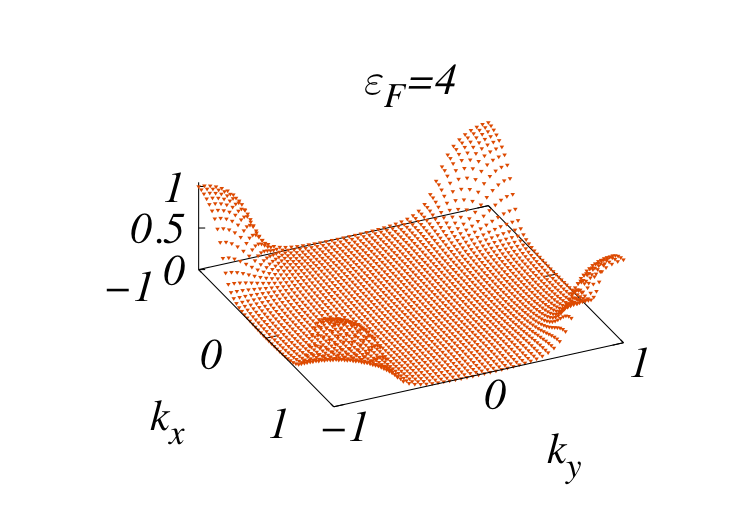}
\caption{\label{fig15}
(Color online) Sum of the absolute square values of the coefficients for the empty state and doubly occupied state (left panels),
for the spin singlet states (middle panels) and triplet states (right panels), for $M_z=1$ and for
$\veps_F=-4,-2,0,2,4$ from top to bottom,
 as a function of momentum.
}
\end{figure*}

For values of the chemical
potential, $|\veps_F|>4$, there is a clear gap between the dominant empty state and all others that is particularly large when
$M_z=1$ (these results are not shown). At the transition to the $\mathbb{Z}_2$ phase the gap closes at some momentum
values.
The absolute value of both triplet components saturates at $1/2$.
This is to be contrasted with the case when there is a finite magnetization. First, in this case the degeneracy
between the two triplet states is lifted. Inside the phase with finite Chern number, the triplet component
$|1100\rangle $ saturates to one at the appropriate singular momentum value (zero momentum in the case of $\veps_F = -4$).
This saturation is characteristic of the phases with finite Chern number.
Increasing the chemical potential, for instance $\veps_F=-2$, the weight of the state $|1111\rangle$ increases
and the weight of the spin triplet state decreases. For $M_z=0$ there is a slight decrease but a clear
difference is noted when $M_z=1$. In the topological phase the triplet state is dominant at some momentum
values while in the trivial phases (with $C=0$) the contribution from other states is also important.
At $\veps_F=0$ ($C=-2$) there is also a saturation of the triplet component, but at different momenta.
It is also clear that, as the chemical potential increases, the weight of the fully occupied state increases, until
it becomes dominant over the Brillouin zone. The roles of the states $|0000\rangle$ and $|1111\rangle$, are naturally reversed
as the chemical potential changes form the bottom to the top of the tight-binding band.

To further understand the relative weights of the basis states we show in Figs. \ref{fig14}, \ref{fig15} the momentum dependence
of the sum of the absolute squares of i) empty and doubly occupied states, ii) singlet pairing states and
iii) triplet pairing states. As mentioned before, these are all the states that give significant contributions.
This allows to better compare the cases $M_z=0$ and $M_z=1$ because the spin orientations become degenerate.
(Also we stress that previous results were presented for the absolute value of the coefficients, while here we
consider the absolute values squared).
In Fig. \ref{fig14} we consider $M_z=0$ and in Fig, \ref{fig15} we consider $M_z=1$.
The various rows correspond to different chemical potentials. In the left column we plot the
sum of the (square of the)  empty and doubly occupied states, in the middle column the singlet pairing
and in the right column the triplet pairing states.

\begin{figure*}[t]
\includegraphics[width=0.19\textwidth]{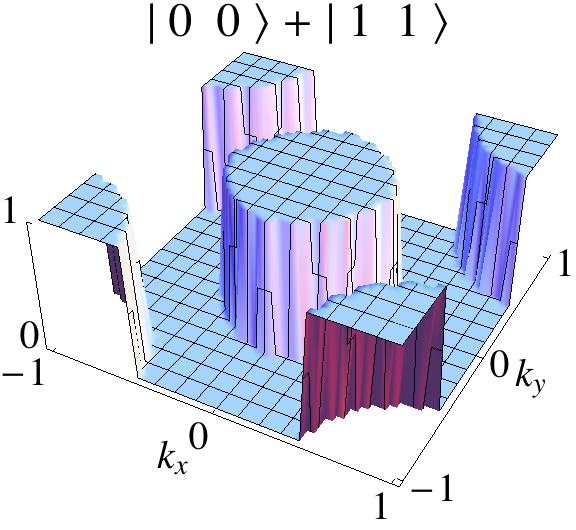}
\includegraphics[width=0.19\textwidth]{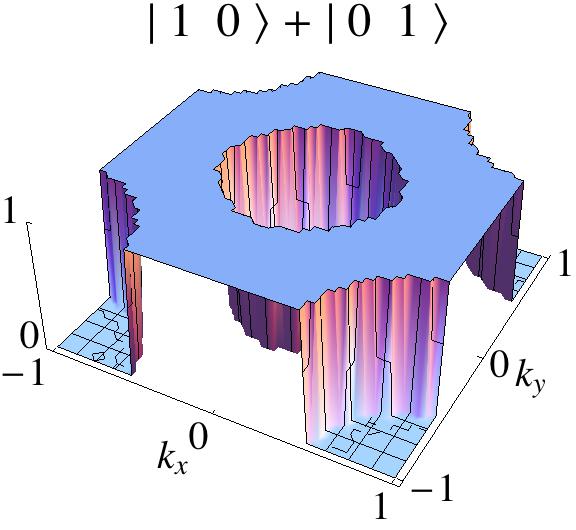}
\includegraphics[width=0.19\textwidth]{fig19a}
\includegraphics[width=0.19\textwidth]{fig19b}
\includegraphics[width=0.17\textwidth]{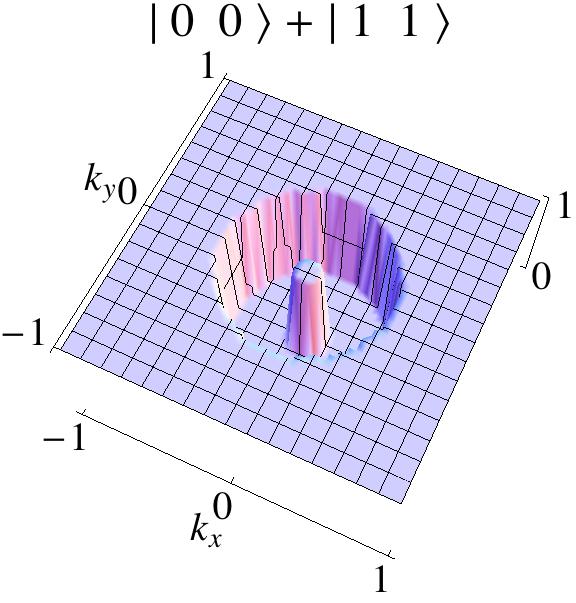}
\caption{\label{fig151}
(Color online) First two panels: Eigenvector components of the highest eigenvalue of the reduced density matrix for
$M_z=2, \veps_F=0$ ($C=-2$).
Following panels: Eigenvector components along the states $|00\rangle$ and $|11\rangle$
of the highest eigenvalue of the reduced density matrix
for $\veps_F=-3$ and $M_z=0.9,1.0,1.1$, respectively.
}
\end{figure*}

The results shown in these figures summarize the various aspects found previously and further clarify
new aspects.
a) The empty and doubly occupied states usually dominate over the Brillouin zone, for all parameter
values, except near the Fermi surface. b) Near the Fermi surface the main contribution comes
from the triplet and singlet pairing states. c) Since we choose the amplitude of the s-wave gap
parameter smaller than the triplet amplitude, the singlet pairing coefficient is smaller than
the triplet. For $M_z=0$ the amplitudes of the singlet and triplet states are similar.
For $M_z=1$ the singlet states contribution is considerably smaller than the triplet one.
d) In the case $M_z=0$ the maximal amplitude of both the singlet and triplet states is quite uniform
along the Fermi surface with small peaks near the singular momenta. e) In the case of finite
magnetization, the triplet coefficient saturates near the singular points. Therefore, the dependence
on momentum along the Fermi surface is stronger. f) As for the entanglement measures, the significance
of the singular momenta is quite clear in the various topological phases.

\section{Eigenvectors of $k$-subspace reduced density matrix}

The $k$-subspace reduced density matrix $\rho^{\prime}$ is defined in a basis of the
occupation numbers $|n_{k,\uparrow},n_{k,\downarrow}\rangle $. The states are therefore
of the type $|00\rangle, |01\rangle, |10\rangle, |11\rangle$. The diagonalization of this
$4\times 4$ matrix leads to the eigenvalues and eigenvectors expressed in this basis.

In the first two panels of Fig. \ref{fig151} we present results for the eigenvector of the largest eigenvalue
for the point in parameter space $M_z=2, \veps_F=0$ ($C=-2$). We plot the sum of the absolute
coefficients along the states $|00\rangle$ and $|11\rangle$, in the top panel and along the
states $|01\rangle$ and $|10\rangle$, in the lower panel. We see that the contributions from
the states with empty or double occupancy and the states with single occupancy, are exclusive
which implies a very sharp separation in the Brillouin zone between the two sets of states.
This rather sharp separation occurs around the phase diagram.

However, a detailed analysis near a transition between the various topological phases
shows that, focusing on the special moments involved in a given transition these provide
a clear signature of the crossing. As an example we consider in the last three panels of Fig. \ref{fig151}
the eigenvector components along the states $|00\rangle$ and $|11\rangle$ of the highest eigenvalue of the reduced density matrix
for $\veps_F=-3$ and $M_z=0.9,1.0,1.1$, respectively.
This transition is associated with the gapless point at momentum $(0,0)$. As Fig. \ref{fig151} shows there is
a clear signature at this momentum value as the transition occurs. In the regime when $C=0$ there is a peak
of the eigenvector components at $k=(0,0)$ which vanishes in the phase with $C=1$,
a typical behavior of the time-reversal momenta at the transition points.
We have checked that in the components of the singly occupied states the same (complementar) feature occurs.


\begin{thebibliography}{}


\bibitem{TKNN}
D. J. Thouless, M. Kohmoto, M. P. Nightingale, and M. den Nijs,
Phys. Rev. Lett. {\bf 49}, 405 (1982).

\bibitem{Hasan}
M. Z. Hasan and C. L. Kane,
 Rev. Mod. Phys. {\bf 82}, 3045 (2010)

\bibitem{Zhang}
 X.-L. Qi and S.-C  Zhang,
 Rev. Mod. Phys.  {\bf 83}, 1057 (2011).

\bibitem{Konig} M. K\"onig, S. Wiedmann, C. Br\"une, A. Roth, H. Buhmann, L. W. Molenkamp, X. L. Qi, and S. C. Zhang,
Science {\bf 318}, 766 (2007).

\bibitem{Hsieh}
D. Hsieh, D. Qian, L. Wray, Y. Xia, Y. S. Hor, R. J. Cava, and M. Z. Hasan,
Nature (London) {\bf 452}, 970 (2008).

\bibitem{Xia}
Y. Xia, D. Qian, D. Hsieh, L. Wray, A. Pal, H. Lin, A. Bansil,
D. Grauer, Y. S. Hor, R. J. Cava, and M. Z. Hasan, Nat.
Phys. {\bf 5}, 398 (2009).


\bibitem{Halperin}
B. I. Halperin, Phys. Rev. B {\bf 25}, 2185 (1982).

\bibitem{Hatsugai}
Y. Hatsugai, Phys. Rev. Lett. {\bf 71}, 3697 (1993).

\bibitem{ReadGreen}
N. Read and D. Green, Phys. Rev. B {\bf 61}, 10267 (2000).

\bibitem{SatoPRL09}
M. Sato, Y. Takahashi, S. Fujimoto, Phys. Rev. Lett. {\bf 103}, 020401 (2009).

\bibitem{SauPRL10}
J. D. Sau, R. M. Lutchyn, S. Tewari, S. Das Sarma, Phys. Rev. Lett. {\bf 104}, 040502 (2010).

\bibitem{Fu}
L. Fu and C. L. Kane, Phys. Rev. Lett. {\bf 100}, 096407 (2008).

\bibitem{QHZ10}
X.-L. Qi, T. L. Hughes, S.-C. Zhang, Phys. Rev. B {\bf 82}, 184516 (2010).

\bibitem{Mourik}
V. Mourik, K. Zuo, S. M. Frolov, S. R. Plissard, E. P.
A. M. Bakkers, and L. P. Kouwenhoven, Science {\bf 336}, 1003 (2012).


\bibitem{Das}
A. Das, Y. Ronen, Y. Most, Y. Oreg, M. Heiblum, and H. Shtrikman, Nat. Phys. {\bf 8}, 887 (2012).


\bibitem{Deng}
M. T. Deng, C. L. Yu, G. Y. Huang, M. Larsson,
P. Caroff, and H. Q. Xu, Nano Letters {\bf 12}, 6414 (2012).

\bibitem{Fan}
Fan Zhang, C.L. Kane and E.J. Mele, arXiv:1212.4232.


\bibitem{Yazdani}
S. Nadj-Perge, I. K. Drozdov, B. A. Bernevig, and Ali Yazdani, arXiv:1303.6363


\bibitem{Veldhorst}
M. Veldhorst, M. Snelder, M. Hoek, T. Gang, V. K. Guduru, X. L. Wang, U. Zeitler,
W. G. van der Wiel, A. A. Golubov, H. Hilgenkamp, and A. Brinkman, Nature Materials {\bf 11}, 417 (2012).


\bibitem{Williams}
J. R. Williams, A. J. Bestwick, P. Gallagher, S. S. Hong, Y. Cui, A. S. Bleich, J. G. Analytis,
I. R. Fisher and D. Goldhaber-Gordon, Phys. Rev. Lett. {\bf 109}, 056803 (2012).


\bibitem{Rokhinson}
L. P. Rokhinson, X. Liu, and J. K. Furdyna, Nature Physics {\bf 8}, 795 (2012).

\bibitem{Pablo}
Pablo San-Jos\'e, Jorge Cayao, Elsa Prada and Ram\'on Aguado, arXiv:1301.4408.

\bibitem{Schnyder} A. P. Schnyder, P. M. R. Brydon and C. Timm, Phys. Rev. B {\bf 85}, 024522 (2012).

\bibitem{Ojanen}
T. Ojanen and T. Kitagawa, Phys. Rev. B {\bf 87}, 014512 (2013).

\bibitem{us3} P.D. Sacramento, M.A.N. Ara\'ujo and E.V. Castro, Europhys. Lett. (2014); arxiv:1302.3122.

\bibitem{amico} L. Amico, R. Fazio, A. Osterloh and V. Vedral, Rev. Mod. Phys. {\bf 80}, 517 (2008).

\bibitem{sachdev} S.~Sachdev, {\em Quantum Phase Transitions}, Cambridge University Press (1999).

\bibitem{latorre-2004-4}
J.I. Latorre, E. Rico and G. Vidal,
Quantum Inf. Comput. {\bf 4}, 48 (2004).

\bibitem{gu-2005-71}
S.-J. Gu, G.-S. Tian and H.-Q. Lin,
Phys. Rev. A {\bf 71}, 052322 (2005).

\bibitem{gu_pra_68}
S.-J. Gu, H.-Q. Lin, and Y.-Q. Li,
Phys. Rev. A {\bf 68}, 042330 (2003).

\bibitem{gu} S.-J. Gu, S.-S. Deng, Y.-Q. Li and H.-Q. Lin, Phys. Rev. Lett.
{\bf 93}, 086402 (2004).

\bibitem{goteborg} D. Larsson and H. Johannesson, Phys. Rev. Lett. {\bf 95}, 196406 (2005).

\bibitem{vidal-2006-73}
J. Vidal,
Phys. Rev. A {\bf 73}, 062318 (2006).

\bibitem{vidal-2006}
J. Vidal, S. Dusuel and T. Barthel,
J. Stat. Mech. {\bf 0701}, 01015 (2007).

\bibitem{gu_prl_93}
S.-J. Gu, S.-S. Deng Y.-Q. Li and H.-Q. Lin,
Phys. Rev. Lett. {\bf 93}, 086402 (2004).

\bibitem{nielsen.chuang}
M.A. Nielsen and I.L. Chuang,
{\bf Quantum Computation and Quantum Information}, Cambridge University Press, 2000.

\bibitem{wooters}
W.K. Wootters,
Phys. Rev. Lett. {\bf 80}, 2245 (1998).

\bibitem{vedral_njp6}
V. Vedral,
New. J. Phys. {\bf 6}, 22 (2004).

\bibitem{gu_qp_1}
S.-J. Gu, C.-P. Sun and H.-Q. Lin,
J. Phys. A {\bf 41}, 025002 (2008).

\bibitem{gu_qp_2}
W.-L. Chan, J.-P. Cao, D. Yang and S.-J. Gu,
J. Phys. A {\bf 40}, 12143 (2007).

\bibitem{vidal} G. Vidal and R.F. Werner, Phys. Rev. A {\bf 65}, 032314 (2002).

\bibitem{meyer}A.D. Meyer and N.R. Wallach, J. Math. Phys. 43, 4273 (2002).

\bibitem{oliveira_pra_73}
T.R. de Oliveira, G. Rigolin and M.C. de Oliveira
Phys. Rev. A {\bf 73}, 010305(R) (2006).

\bibitem{miranda} T.R. de Oliveira, G. Rigolin, M.C. de Oliveira and E. Miranda, Phys. Rev. Lett.
{\bf 97}, 170401 (2006).

\bibitem{lunkes05-2}
C. Lunkes, \v{C}. Brukner and V. Vedral,
Phys. Rev. Lett. {\bf 95}, 030503 (2005).

\bibitem{heaney06}
L. Heaney, J. Anders and V. Vedral,
quant-ph/0607069.

\bibitem{oh04}
S. Oh and J. Kim,
Phys. Rev. B {\bf 71}, 144523 (2005).

\bibitem{Gu} S.J. Gu, Int. J. Mod. Phys. B {\bf 24}, 4371 (2010).

\bibitem{zanardi-first} P. Zanardi and N. Paunkovi\'{c}, Phys. Rev. E {\bf 74}, 031123 (2006).

\bibitem{zanardi-free_fermion} P. Zanardi, M. Cozzini and P. Giorda,  J. Stat. Mech.: Theory Exp. (2007), L02002; M. Cozzini, P. Giorda and P. Zanardi, Phys. Rev. B {\bf 75}, 014439 (2007). 

\bibitem{buonsante-prl} P. Buonsante and A. Vezzani, Phys. Rev. Lett. {\bf 98}, 110601 (2007).

\bibitem{oelkers} N. Oelkers and J. Links, Phys. Rev. B {\bf 75}, 115119 (2007).

\bibitem{chen-excited} S. Chen, L. Wang, S.-J. Gu and Y. Wang, Phys. Rev. E {\bf 76}, 061108 (2007). 

\bibitem{min} M.-F. Yang, Phys. Rev. B {\bf 76}, 180403(R) (2007). 

\bibitem{zhou} H.-Q. Zhou and J. P. Barjaktarevic, e-print arXiv:cond-mat/0701608;
H.-Q. Zhou, J.-H. Zhao and B. Li, J. Phys. A {\bf 41}, 492002 (2008).

\bibitem{zanardi-differential} P. Zanardi, P. Giorda and M. Cozzini, Phys. Rev. Lett. {\bf 99}, 100603 (2007). 

\bibitem{zanardi-scaling} L. CamposVenuti, and P. Zanardi, Phys. Rev. Lett. {\bf 99}, 095701 (2007). 

\bibitem{wen-long-thermal} W.-L. You, Y.-W. Li and S.-J. Gu, Phys. Rev. E {\bf 76}, 022101 (2007). 

\bibitem{Zhou} H.Q. Zhou, arxiv:0704.2945.

\bibitem{us} N. Paunkovi\'c, P.D. Sacramento, P. Nogueira, V.R. Vieira and V.K. Dugaev,
Phys. Rev. A {\bf 77}, 052302 (2008).

\bibitem{BCS} N. Paunkovi\'c and V.R. Vieira, Phys. Rev. E {\bf 77}, 011129 (2008).

\bibitem{Zanardi_t1} P. Zanardi, H. T. Quan, X. Wang and. C. P. Sun, Phys. Rev. A {\bf 75}, 032109 (2007).

\bibitem{Zanardi_t2} P. Zanardi, L. C. Venuti and P. Giorda, Phys. Rev. A {\bf 76}, 062318 (2007).


\bibitem{Haldane} H. Li and F.D.M. Haldane, Phys. Rev. Lett. {\bf 101}, 010504 (2008).

\bibitem{Bernevig} N. Regnault, B. A. Bernevig, and F. D. M. Haldane, Phys. Rev. Lett.
{\bf 103}, 016801 (2009).

\bibitem{Poilblanc} D. Poilblanc, Phys. Rev. Lett. {\bf 105}, 077202 (2010).

\bibitem{Bernevig2} A. Sterdyniak, N. Regnault, and B. A. Bernevig,
Phys. Rev. Lett. {\bf 106}, 100405 (2011).

\bibitem{Arovas} R. Thomale, D.P. Arovas and B.A. Bernevig, Phys. Rev. Lett. {\bf 105},
116805 (2010).

\bibitem{us2} P.D. Sacramento, N. Paunkovic and V.R. Vieira, Phys. Rev. A {\bf 84}, 062318 (2011).

\bibitem{guafter} S.-J. Gu, W.-C. Yu and H.-Q. Lin, arXiv:1108.2832.


\bibitem{chen77} S. Chen, L. Wang, Y. Hao and Y. Wang, Phys. Rev. A {\bf 77}, 032111 (2008).

\bibitem{gu77} S.-J. Gu, H.-M. Kwok, W.-Q. Ning and H.-Q. Lin, Phys. Rev. B {\bf 77}, 245109 (2008).

\bibitem{hamma77} A. Hamma, W. Zhang, S. Haas and D.A. Lidar, Phys. Rev. B {\bf 77}, 155111 (2008).

\bibitem{abasto78} D.F. Abasto, A. Hamma and P. Zanardi, Phys. Rev. A {\bf 78}, 010301 (2008).

\bibitem{yang78} S. Yang, S.-J. Gu, C.-P. Sun and H.-Q. Lin, Phys. Rev. A {\bf 78}, 012304 (2008).

\bibitem{campus78} L. Campus Venuti, M. Cozzini, P. Buonsante, F. Massel, N. Bray-Ali and P. Zanardi,
Phys. Rev. B {\bf 78}, 115410 (2008).

\bibitem{abasto79} D.F. Abasto and P. Zanardi, Phys. Rev. A {\bf 79}, 012321 (2009).

\bibitem{zhao80} J.-H. Zhao and H.-Q. Zhou, Phys. Rev. B {\bf 80}, 014403 (2009).

\bibitem{eriksson} E. Eriksson and H. Johannesson, Phys. Rev. A {\bf 79}, 060301(R) (2009).

\bibitem{castelnovo} C. Castelnovo and C. Chamon, Phys. Rev. B {\bf 77}, 054433 (2008).

\bibitem{trebst} S. Trebst et al., Phys. Rev. Lett. {\bf 98}, 070602 (2007).

\bibitem{wang10} Z. Wang, T. Ma, S.-J. Gu and H.-Q. Lin, Phys. Rev. A {\bf 81}, 062350 (2010).

\bibitem{wang2013} Z. Wang, Q.F. Liang and D.X. Yao, arXiv:1302.5492.

\bibitem{Gorkov}
L. P. Gor'kov and E. I. Rashba, Phys. Rev. Lett. {\bf 87}, 037004 (2001).

\bibitem{sato}
M. Sato and S. Fujimoto, Phys. Rev. B {\bf 79}, 094504 (2009).

\bibitem{Sigrist2}
P. A. Frigeri, D. F. Agterberg, A. Koga and M. Sigrist, Phys. Rev. Lett. {\bf 92}, 097001 (2004).

\bibitem{ahe} P.D. Sacramento, M.A.N. Ara\'ujo, V.R. Vieira, V.K. Dugaev and J. Barnas
Phys. Rev. B {\bf 85}, 014518 (2012).

\bibitem{xiao}
D. Xiao, M.-C. Chang and Q. Niu, Rev. Mod. Phys. {\bf 82}, 1959 (2010).

\bibitem{Fukui}
T. Fukui, Y. Hatsugai and H. Suzuki, J. Phys. Soc. Jpn. {\bf 74}, 1674 (2005).

\bibitem{Ludwig} A.P. Schnyder, S. Ryu, A. Furusaki and A.W.W. Ludwig, Phys. Rev. B {\bf 78},
195125 (2008)

\bibitem{LudwigAIP} A.P. Schnyder, S. Ryu, A. Furusaki and A.W.W. Ludwig,
in {\it Advances in Theoretical Physics}, edited by Vladimir Lebedev and Mikhail Feigel'man,
AIP Conf. Proc. No. 1134 (AIP, Melville, NY, 2009), p. 10.

\bibitem{LudwigNJP} S. Ryu, A.P. Schnyder, A. Furusaki and A.W.W. Ludwig,
New J. Phys. {\bf 12}, 065010 (2010).

\bibitem{Wen} X. G. Wen and A. Zee, Nucl. Phys. B {\bf 316}, 641 (1989).

\bibitem{Jozsa} R. Jozsa, J. Mod. Opt. {\bf 41}, 2315 (1994).








\end{thebibliography}
\end{document}